\documentclass[pdflatex,sn-mathphys-num]{sn-jnl}% Math and Physical Sciences Numbered Reference Style
\usepackage{graphicx}%
\usepackage{multirow}%
\usepackage{amsmath,amssymb,amsfonts}%
\usepackage{amsthm}%
\usepackage{mathrsfs}%
\usepackage[title]{appendix}%
\usepackage{xcolor}%
\usepackage{textcomp}%
\usepackage{manyfoot}%
\usepackage{booktabs}%
\usepackage{algorithm}%
\usepackage{algorithmicx}%
\usepackage{algpseudocode}%
\usepackage{listings}%
\usepackage{makecell}%
\usepackage{seqsplit}%
\usepackage{placeins}
\usepackage{bibunits}
\theoremstyle{thmstyleone}%
\theoremstyle{thmstyletwo}%

\theoremstyle{thmstylethree}%

\begin{document}

\title[Article Title]{Cosm: orchestrated collective dynamics give rise to rapid computational search at the solution frontier}


\author*[1]{\fnm{Kenneth M.} \sur{Zick}}\email{kzick@isi.edu}

\author[2]{\fnm{Nikhil} \sur{Shukla}}

\author[3]{\fnm{Alexander} \sur{Marakov}}

\affil*[1]{\orgname{University of Southern California, Information Sciences Institute}, \orgaddress{\city{Arlington}, \state{Virginia}, \country{USA}}}

\affil[2]{\orgname{University of Virginia}, \orgaddress{\city{Charlottesville}, \state{Virginia}, \country{USA}}}

\affil[3]{\orgname{ Northrop Grumman Systems Corporation}, \orgaddress{\city{Linthicum}, \state{Maryland}, \country{USA}}}


\abstract{We introduce Collective Switched Motion (Cosm), a heuristic optimization framework based on orchestrated collective dynamics. Cosm organizes the objective into conflict-minimized subobjectives and activates them sequentially, temporally separating competing local influences. The interplay of switching and local interactions among variables, supplemented by structured correlated perturbations, gives rise to collective search behavior including coordinated cluster motion. Using this framework, we develop and test a solver for sparse Ising optimization. On G64, a large sparse instance with a highly heterogeneous degree distribution from the longstanding Gset Max-Cut benchmark suite, Cosm establishes a new best-known cut of 8753. On large random-graph instances G61 and G70, whose best-known cuts of 5799 and 9595 have been difficult for prior heuristics to attain, a vectorized CPU implementation of Cosm reliably reaches these cuts with times-to-target of 15 s and 2.4 s, respectively. Cosm also heuristically attains the certified optimum of G81, the largest Gset instance (a 20,000-variable two-dimensional spin glass), along with those of G72 and G77. Tests on planted-solution spin-glass benchmarks suggest more favorable empirical scaling than that previously reported for seven compared solvers. Cosm demonstrates that, from simple interacting elements and a new form of orchestration, effective search can emerge.}




\maketitle

\section{Introduction}

Searching an enormous space for a superior solution is a fundamental computational task arising in optimization, computational science, and decision-making. The solution could represent a safety-critical trajectory, a low-energy structure of a molecule or material, or a strategy for a high-stakes decision. With large numbers of variables and competing influences, the search landscape can be extraordinarily rugged and complex, and reaching or advancing the practical frontier of solution quality can require prohibitive computational time and energy.

Many conventional approaches to heuristic search begin by explicitly designing procedures for navigating the solution landscape. Dynamical approaches open another possibility: useful search mechanisms may be found in the collective behavior of interacting elements, without the resulting search trajectory being explicitly prescribed. Because the system-level consequences of even simple interacting dynamics can be difficult to anticipate, the large space of possible collective dynamics represents a potential discovery space for computational mechanisms. A key question is whether such mechanisms, whether deliberately designed or discovered through exploration, can be organized and combined to produce effective search.

An important class of optimization problems involves binary variables coupled through highly sparse pairwise interactions. These sparse Ising-type problems, including Max-Cut and quadratic unconstrained binary optimization (QUBO) problems, arise in applications including task scheduling, spin-glass simulations, and power systems optimization \cite{schuetz2022,wurtz2024,barrass2025}. Sparse local interaction structures are commonplace due to physical, communication, or architectural constraints. These problems often give rise to rugged landscapes and frustration that challenge exact solvers~\cite{charfreitag2022,rehfeldt2023} and semidefinite relaxation methods \cite{choi2000}. Sparse Ising-type problems have motivated a broad range of optimization approaches, including Monte Carlo heuristics such as simulated annealing and parallel tempering~\cite{zhu2015}, learning-based approaches~\cite{fan2023}, quantum annealing~\cite{tasseff2024,munoz2025}, and classical dynamical-system heuristics. Among the latter, the Simulated Bifurcation Machine (SBM) introduced by Goto et al.~\cite{goto2021} has demonstrated state-of-the-art performance on dense problem instances~\cite{mohseni2022,goto2026}. However, SBM and several other dynamical approaches, such as the Coherent Ising Machine (CIM), have exhibited lower performance on sparse instances, as seen in a recent comparative study by Hou et al.~\cite{hou2025}.

Ising spin glasses provide a particularly demanding setting for heuristic search. They are disordered and frustrated magnetic systems with highly rugged energy landscapes. Many hard combinatorial optimization problems can be formulated as Ising models and likewise exhibit frustration and ruggedness. This connection has helped establish spin glasses as canonical benchmarks for optimization solvers and Ising machines~\cite{katzgraber2018}. Upon surveying the field of spin-glass dynamics in 2025, Parisi and coauthors report that, beyond parallel tempering, algorithmic progress has been largely stagnant over the past two decades and that new algorithms are needed for both physics and engineering applications~\cite{dahlberg2025}.

These observations motivate exploring alternative search dynamics. In this work, we introduce Collective Switched Motion (Cosm), a heuristic optimization framework based on orchestrated collective dynamics. The main contributions are summarized as follows:
\begin{itemize}
\item A heuristic framework that organizes an objective into conflict-minimized subobjectives and activates them sequentially, allowing collective search behavior to emerge from the orchestrated dynamics.
\item Empirical characterization across planted-solution spin-glass instances and large benchmark problems, including a lower empirical scaling exponent than previously reported for seven compared solvers, attainment of certified optima on large Gset problems and discovery of a new best-known solution on G64, an instance with a broad degree distribution.
\item A vectorized CPU implementation demonstrating that Cosm's collective search dynamics can be implemented efficiently using regular, branchless integer computation and SIMD parallelism, with 99\%-probability times-to-target measured in seconds on G60, G61, and G70 and approximately three minutes on G64.
\end{itemize}

\section{Cosm framework}
\subsection{Compilation of objective into a switched dynamical system}

Many established search heuristics combine the contributions of multiple objective function terms when determining how the search state should evolve, for example through gradients, local fields, or changes in objective value associated with candidate moves. In dynamical optimization, for example, local interactions are commonly combined into a net field or force that governs the evolution of each variable. Competing local influences can partially or exactly cancel when combined into an aggregate update, potentially producing a weak or vanishing net effect even when the individual contributions are substantial. Cosm takes a different approach. It compiles the objective into a set of subobjectives constructed to reduce or eliminate simultaneous conflicts among local interactions. Search proceeds iteratively by repeatedly traversing the set of subobjectives, activating them sequentially to temporally separate competing local influences. The search trajectory emerges from the resulting switched dynamics.

The objective function $f(\mathbf{x})$, where $\mathbf{x}$ is a vector of $n$ variables, is decomposed into $C$ subobjectives according to
\begin{equation}
f(\mathbf{x}) = \sum_{c=1}^{C} f_c(\mathbf{x})
\end{equation}
where $C$ depends on the chosen decomposition. The subobjectives are constructed to reduce or eliminate conflicts within each $f_c$, where a conflict occurs when objective terms sharing a variable can exert competing influences on its update. In the Ising realization of Cosm in Section~\ref{sec: cosm_for_Ising}, edge coloring eliminates such conflicts entirely: each variable participates in at most one pairwise interaction per subobjective. 

The Cosm update rule sequentially applies the dynamics associated with individual subobjectives. A complete traversal of all $C$ subobjectives constitutes a \emph{sweep}, while each activation constitutes a \emph{subsweep}. Specifically, at subsweep $t$, the update is given by
\begin{equation}
\label{eq:update_rule}
\mathbf{x}_{t+1}=\Phi_{c(t)}(\mathbf{x}_t; \alpha_t)
\end{equation}
where $\Phi_c$ is the dynamical update associated with subobjective $f_c$ and $\alpha_t$ is an update scale. Within each sweep, the subobjective activation order is specified by a switching policy $c(t)$. The ordering may be fixed across sweeps, dynamically permuted, or selected according to a state-dependent rule, provided that each subobjective is activated once per sweep. 

Unlike variable-block approaches, Cosm decomposes the objective rather than the variable set. Each subobjective need not be restricted to a small set of variables, and indeed can contain all variables. Moreover, the decomposition is not merely a means of simplifying or parallelizing evaluation; it determines how local interactions are organized in time to generate the search dynamics.

\subsection{Switched collective dynamics}

In Cosm, variables interact according to simple local rules specified by the active subobjective. Sequential activation couples these local responses across time, giving rise to a particular form of orchestrated collective dynamics that we refer to as switched collective dynamics. For $t$ corresponding to the beginning of a sweep, and with the constant within-sweep update scale left implicit, a single sweep consists of the composition
\begin{equation}
\mathbf{x}_{t+C}=(\Phi_{c_{C}} \circ \cdots \circ \Phi_{c_{1}})(\mathbf{x}_t)
\end{equation}
where $(c_1,\ldots,c_C)$ denotes the subobjective activation order for that sweep.
Let $\Phi_\mathrm{all}$ denote a corresponding one-sweep update in which the influences of the subobjectives are aggregated rather than applied sequentially. In general, 
\begin{equation}
(\Phi_{c_{C}} \circ \cdots \circ \Phi_{c_{1}})(\mathbf{x}_t) \neq \Phi_\mathrm{all}(\mathbf{x}_t)
\end{equation}
because each subobjective acts on a state already modified by the preceding subobjectives.

Although each subobjective consists of conflict-minimized local interactions, successive subobjectives may couple overlapping sets of variables. Motion induced during one subsweep may therefore alter the state encountered by subsequent interactions, allowing coordinated behavior to develop across the interaction network.

\subsection{Update scale}
\label{sec: update scale}
The magnitude of the local dynamics can be regulated by a global update scale $\alpha_t$. The scale may vary over the course of the search according to an annealing or other schedule. Because each subsweep acts on the state produced by preceding subsweeps, the update scale can affect the resulting composed dynamics rather than simply the rate at which a fixed trajectory is traversed. Varying the update scale can therefore allow the system to traverse different dynamical regimes as the search proceeds.

\subsection{Structured correlated perturbation}
A further design principle of Cosm is the use of structured correlated perturbations to complement the switched dynamics and facilitate coordinated transitions between configurations. Rather than perturbing variables independently, correlated perturbations select sets of variables and perturb them coherently, allowing collective motion while approximately preserving selected relationships within the perturbed set. Selection may depend on the current state $\mathbf{x}_t$, problem structure, stochastic rules, or other criteria. The selection rule, perturbation transformation, timing, and magnitude are design choices that may vary across realizations of the framework. At each subsweep, a perturbation decision $q_t\in\{0,1\}$ determines whether the perturbation function $\mathcal{P}_t$ is applied:
\begin{equation}
\label{eq: structured_perturbation}
\mathbf{x}_{t+1} = \begin{cases}\Phi_{c(t)}(\mathbf{x}_t; \alpha_t), & q_t=0, \\
\mathcal{P}_t(\Phi_{c(t)}(\mathbf{x}_t; \alpha_t)), & q_t=1.
\end{cases}
\end{equation} 

The timing of perturbations may follow a predetermined schedule, be stochastic, or depend on the evolving system state. The perturbation magnitude may be fixed or vary according to a predetermined or state-dependent rule, including coupling to the global update scale.

\section{Cosm for sparse Ising optimization}
\label{sec: cosm_for_Ising}
We next develop a sparse Ising realization of the general Cosm framework. Table~\ref{tab: Ising realization of framework} summarizes how the principal elements of the framework are realized in the Ising solver developed here.

\begin{table}[h]
\centering
\caption{Principal elements of the Cosm framework and their Ising realization.}
\label{tab: Ising realization of framework}
% Resize implies we force it to textwidth, ensuring it fits perfectly.
%\resizebox{\textwidth}{!}{%
%{
\begin{tabular}{l l}
\toprule
  \textbf{Cosm framework concept} &
  \textbf{Cosm Ising realization} \\
\midrule
Objective decomposition & Proper edge coloring into matching-based subobjectives \\
Switching policy & Fixed cyclic activation of edge colors \\
Variable state & Continuous circular variables \\
Local dynamics & Sign-based pairwise interactions \\
Global update scale & Step size $\alpha_t$, linearly annealed \\
Correlated perturbation & Dual window twist (DWT) \\
Readout & Bisection of circular variables into binary spins \\
\bottomrule
\end{tabular}%
\end{table}

\subsection{Objective compilation}
The Ising Hamiltonian is given by
\begin{equation}
H(\mathbf{s}) = -\sum_{(i,j) \in E_G} J_{ij}s_i s_j - \sum_i h_i s_i
\label{eq: IsingHam}
\end{equation}
with $n$ spin variables $s_i \in \{-1, +1\}$, edges ($i$,$j$) in the set $E_G$, pairwise interaction weights $J_{ij}$, and local bias fields $h_i$. For the Ising realization considered here, conflicts within each subobjective are eliminated entirely using a proper $C$-color edge coloring of the input graph $G$. The edge set $E_G$ is partitioned as
\begin{equation}
E_G = E_1 \cup E_2 \cup \cdots \cup E_C,
\end{equation}
where each subset $E_c$ corresponds to a matching associated with color $c$. By definition, no two edges in a matching share a common vertex, and thus no variable appears in more than one term of a given subobjective. A minimum coloring is not required; any proper edge coloring produces conflict-free subobjectives.

The objective function is correspondingly decomposed as
\begin{equation}
f(\mathbf{s}) = \sum_{c=1}^{C} f_c(\mathbf{s}),
\end{equation}
where, for the problems considered here, which have no local bias fields,
\begin{equation}
f_c(\mathbf{s}) = -\sum_{(i,j) \in E_c} J_{ij} s_i s_j.
\end{equation}

\subsection{Continuous relaxation and circular variables}
We next relax the binary spin variables $s_i$ by replacing each with a continuous variable $x_i$ on the unit circle (i.e., a 1-sphere $S^1$), where $x_{i} \in [0, 360)$ degrees with $0\sim360$ (wrapped interval). Updates use modular arithmetic on $[0, 360)$. Circular variables have been employed in approaches such as oscillator Ising machines (OIM)~\cite{wang2019,csaba2020,ZickPatent2023}, XY machines~\cite{ouyang2024}, and semidefinite programming. 

\subsection{Variable interactions}
\begin{figure}
\centering
\includegraphics[width=0.5\columnwidth]{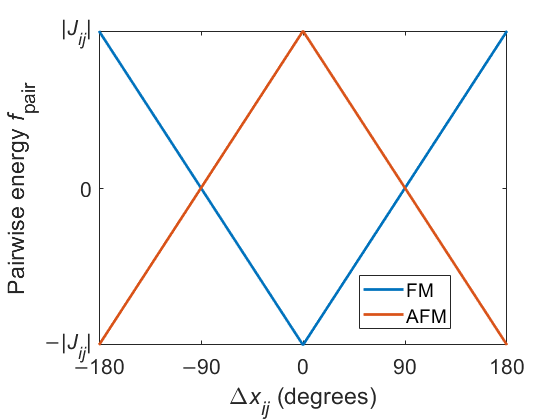}
\caption{Pairwise coupling energy $f_{\mathrm{pair}}(x_i,x_j)=-J_{ij}v(x_i,x_j)$ for ferromagnetic (FM) and antiferromagnetic (AFM) interactions, using the piecewise-linear coupling function $v(x_i,x_j)$ defined in equation~(\ref{eq:coupling_function}). Energy is minimized for aligned FM variables and diametrically opposed AFM variables.}
\label{fig:pairwiseEnergy}
\end{figure}

We replace the quadratic interaction term $s_i s_j$ with a relaxed coupling between circular variables. Intuitively, we seek a piecewise-linear pairwise energy potential that depends on the angular separation of the variables and is V-shaped for ferromagnetic couplings and inverted-V-shaped for antiferromagnetic couplings. The energy is minimized when an FM (AFM) coupled pair is aligned (diametrically opposed) on the unit circle. We define the wrapped angular difference between $x_i$ and $x_j$ as
\begin{equation}
\label{eq:wrapped_angular_difference}
\Delta x_{ij} = ((x_i-x_j+180^\circ) \bmod 360^\circ)-180^\circ 
\end{equation} 
where mod returns a nonnegative remainder. We then define the coupling function
\begin{equation}
\label{eq:coupling_function}
v(x_i, x_j) = 1-\frac{|\Delta x_{ij}|}{90^\circ} 
\end{equation} 
which takes values in the continuous range $[-1,1]$. This gives the pairwise coupling energy
\begin{equation}
\label{eq:pairwise_energy_function}
f_\mathrm{pair}(x_i,x_j)=-J_{ij}v(x_i,x_j)
\end{equation}
as illustrated in Fig.~\ref{fig:pairwiseEnergy}.

We then define the effect of a pairwise interaction on the two coupled variables. The variables are updated in the direction opposite to the selected partial derivatives of the pairwise energy. While $f_\mathrm{pair}$ is not differentiable at angular separations of $0$ or $\pm180$, we define a biased signum (sgn) function for the wrapped angular difference:
\begin{equation}
\label{eq: biased_sgn_function}
\operatorname{sgn_b}(\Delta x_{ij}) = \begin{cases}1, & \Delta x_{ij} \geq 0, \\
-1, & \Delta x_{ij} < 0.
\end{cases}
\end{equation} 
We define gradient-like pairwise update components using the following selected partial derivatives:
\begin{equation}
\label{eq:selected_derivate}
\begin{aligned}
\frac{\widetilde\partial f_\mathrm{pair}}{\partial x_i} &= -J_{ij}\frac{(-\operatorname{sgn_b}(\Delta x_{ij}))}{90^\circ}, \\
\frac{\widetilde\partial f_\mathrm{pair}}{\partial x_j} &= -J_{ij}\frac{(\operatorname{sgn_b}(\Delta x_{ij}) )}{90^\circ}.
\end{aligned}
\end{equation}
We define a normalized gradient-like vector \(\widetilde{\nabla}f\) with respect to the variables in $\mathbf{x}$ by multiplying the selected partial derivatives by \(90^\circ\), allowing the global update scale \(\alpha_t\) to control angular step sizes in degrees. The resulting pairwise variable updates are
\begin{equation}
\label{eq: pair_interaction_contributions}
\begin{aligned}
\Delta x_i &=-\alpha_t \left( 90^\circ\frac{\widetilde\partial f_\mathrm{pair}}{\partial x_i} \right) =-\alpha_t J_{ij}\operatorname{sgn_b}(\Delta x_{ij}), \\
\Delta x_j &=-\alpha_t \left(90^\circ\frac{\widetilde\partial f_\mathrm{pair}}{\partial x_j}\right) =\alpha_t J_{ij}\operatorname{sgn_b}(\Delta x_{ij}).
\end{aligned}
\end{equation} 
Note that unlike sinusoidal interactions, the interactions here maintain update magnitudes independent of angular separation and do not weaken continuously with angular distance. For a given coupling weight and step size, equations~(\ref{eq: biased_sgn_function}) and~(\ref{eq: pair_interaction_contributions}) yield only two possible pairwise updates, analogous to the extremal actions of a bang-bang control policy. Interactions of this type have been used in recent works such as Refs.~\cite{ZickPatent2023,sreedhara2024,mazumderV2_2025}. 

Without Cosm's objective decomposition, a relaxed aggregate objective would be
\begin{equation}
\label{eq: relaxed_objective_function}
f_\mathrm{all}(\mathbf{x}) = -\sum_{(i,j) \in E_G} J_{ij}v(x_i, x_j) = \sum_{(i,j) \in E_G} f_\mathrm{pair}(x_i,x_j)
\end{equation}
with local bias fields omitted for clarity. In this context, a traditional gradient-descent-like approach would employ an update rule of the form
\begin{equation}
\label{eq:aggregate_update}
\mathbf{x}_{t+1}=\mathbf{x}_t-\alpha_t \widetilde\nabla f_\mathrm{all}(\mathbf{x}_t).
\end{equation}
We find that with circular variables and the model described above, using the aggregate update does not provide competitive accuracy on the sparse problems we test. Absent a formal proof, we provide an intuition for one of the difficulties. Unlike the non-periodic variables used in many methods such as SBM, circular variables have two alternative paths (clockwise and counterclockwise) to move around the circle. With aggregate updates, a variable moves by an amount that is the sum over all of its interactions. If the contributions are in opposite directions, the sum can approach zero and the variable can remain stationary, even if a lower energy configuration were possible by rotating 180 degrees. The scenario is loosely analogous to a coordination failure in multi-agent systems, in which agents share a goal but do not achieve a consensus on which of two directions to pursue.
This is a particular type of suboptimal, metastable state we refer to as a \emph{circular conflict trap}. A minimal example in Fig.~\ref{fig:aggregate_vs_SCS}a illustrates how, given equations~(\ref{eq: pair_interaction_contributions})--(\ref{eq:aggregate_update}), even a very small system can become trapped. No binary partition of the frozen configuration yields a ground state. While this example is illustrative, similar cancellation effects can arise in larger sparse graphs due to competing local interactions.

\subsection{Sequential conflict-free search}
\begin{figure}
\centering
\includegraphics[width=0.5\textwidth]{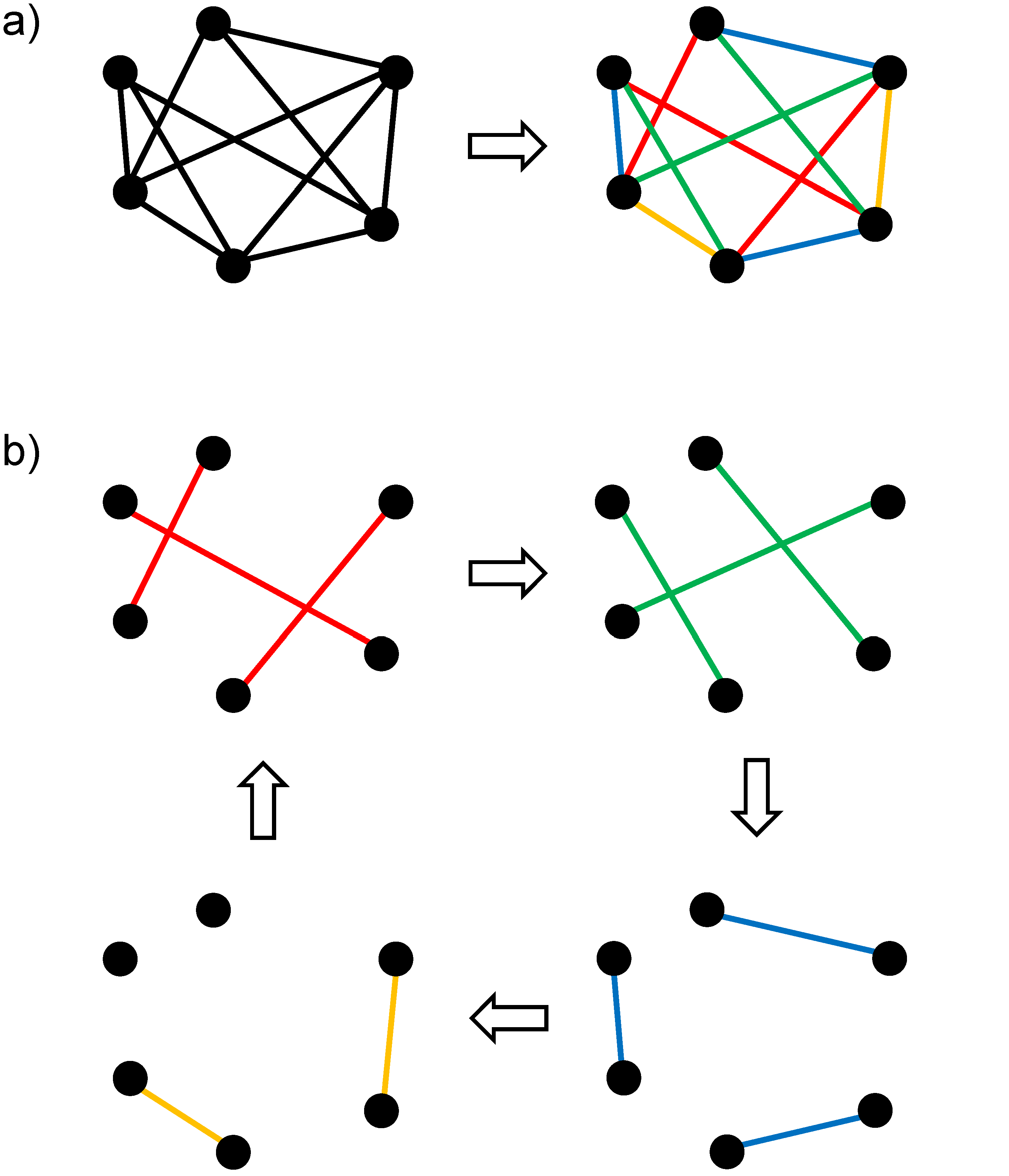}
\caption{Edge coloring and a periodically switched network. a) A proper edge coloring is assigned to the input graph, either by loading a precomputed coloring or generating one using a polynomial-time heuristic. b) The edge-colored graph is partitioned by color into \emph{C} matchings (each with maximum degree one), ensuring that simultaneously active interactions are conflict-free. The network switches periodically among the conflict-free matchings.}
\label{fig:edgeColoring}
\end{figure}

Rather than updating the system using the aggregate derivative vector $\tilde\nabla f_\mathrm{all}(\mathbf{x})$, the Cosm realization studied here sequentially applies updates associated with the individual subobjectives $f_c$. We use a cyclic switching schedule, with the active subobjective index at subsweep $t$ given by
\begin{equation}
\label{eq: active_subobjective_index}
c(t) = ((t-1) \bmod C) + 1.
\end{equation}
The resulting color sequence is $1,2,\ldots,C,1,2,\ldots,C,\ldots$, with each cycle constituting a sweep. The corresponding subsweep update is given by
\begin{equation}
\label{eq:Ising_realization_update_rule}
\mathbf{x}_{t+1}=\mathbf{x}_t-\alpha_t \widetilde\nabla f_{c(t)}(\mathbf{x}_t).
\end{equation}
Effectively, the system evolves on an interaction network that is rhythmically reconfigured as the edge partitions are activated in sequence. This is a particular type of switched (time-varying) network we refer to as a \emph{periodically switched network}. During each subsweep, a variable interacts with at most one other variable, so simultaneously active interactions are conflict-free. An example of such a network is illustrated in Fig.~\ref{fig:edgeColoring}. We refer to the resulting dynamical mechanism as \emph{sequential conflict-free search} (SCS).  

\begin{figure}[h]
\centering
\includegraphics[width=\columnwidth]{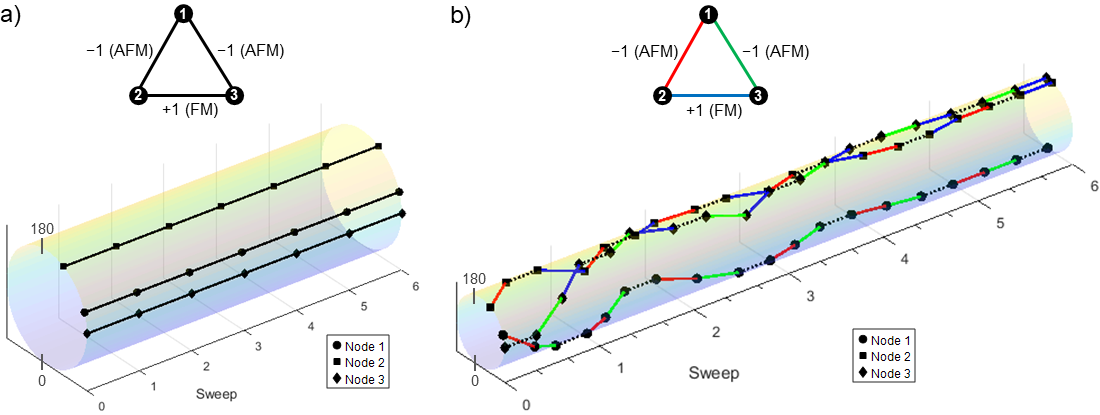}
\caption{Comparison of aggregate-update dynamics and sequential conflict-free search (SCS). a) Top: Seemingly simple three-variable Ising problem with no frustration and two ground states (up-down-down and down-up-up, Ising energy $-3$). Bottom: Trajectory of the circular-variable system under aggregate updates. From the particular random initial state shown, the aggregate contributions cancel, yielding $\widetilde{\nabla}f_{\mathrm{all}}=0$ and frozen dynamics at a suboptimal configuration (Ising energy $-1$). b) Top: Three-variable problem instance after edge coloring. Bottom: System trajectory as the three edge-color partitions are activated in sequence. Trajectory segments are colored according to the active edge-color partition; dotted black segments indicate subsweeps in which a variable is inactive and maintains its state. Unlike in a), the system escapes from the circular conflict trap and reaches a configuration corresponding to a ground state within one sweep (Ising energy $-3$).}
\label{fig:aggregate_vs_SCS}
\end{figure}

SCS leads to \emph{intra-sweep variable fluctuations} in which variable movements (phase rotations) accumulate from subsweep to subsweep. These alter the search dynamics in ways that can empirically improve performance. A minimal example of SCS dynamics and fluctuations is shown in Fig.~\ref{fig:aggregate_vs_SCS}b; unlike the baseline aggregate-update case (Fig.~\ref{fig:aggregate_vs_SCS}a), the system avoids a circular conflict trap and reaches an optimal configuration. Intra-sweep fluctuations are also visible in the larger Cosm dynamical system example shown in Fig.~\ref{fig:varTrajectory1}b.

\subsection{Dual window twist correlated perturbation}

\begin{figure}
\centering
\includegraphics[width=0.6\columnwidth]{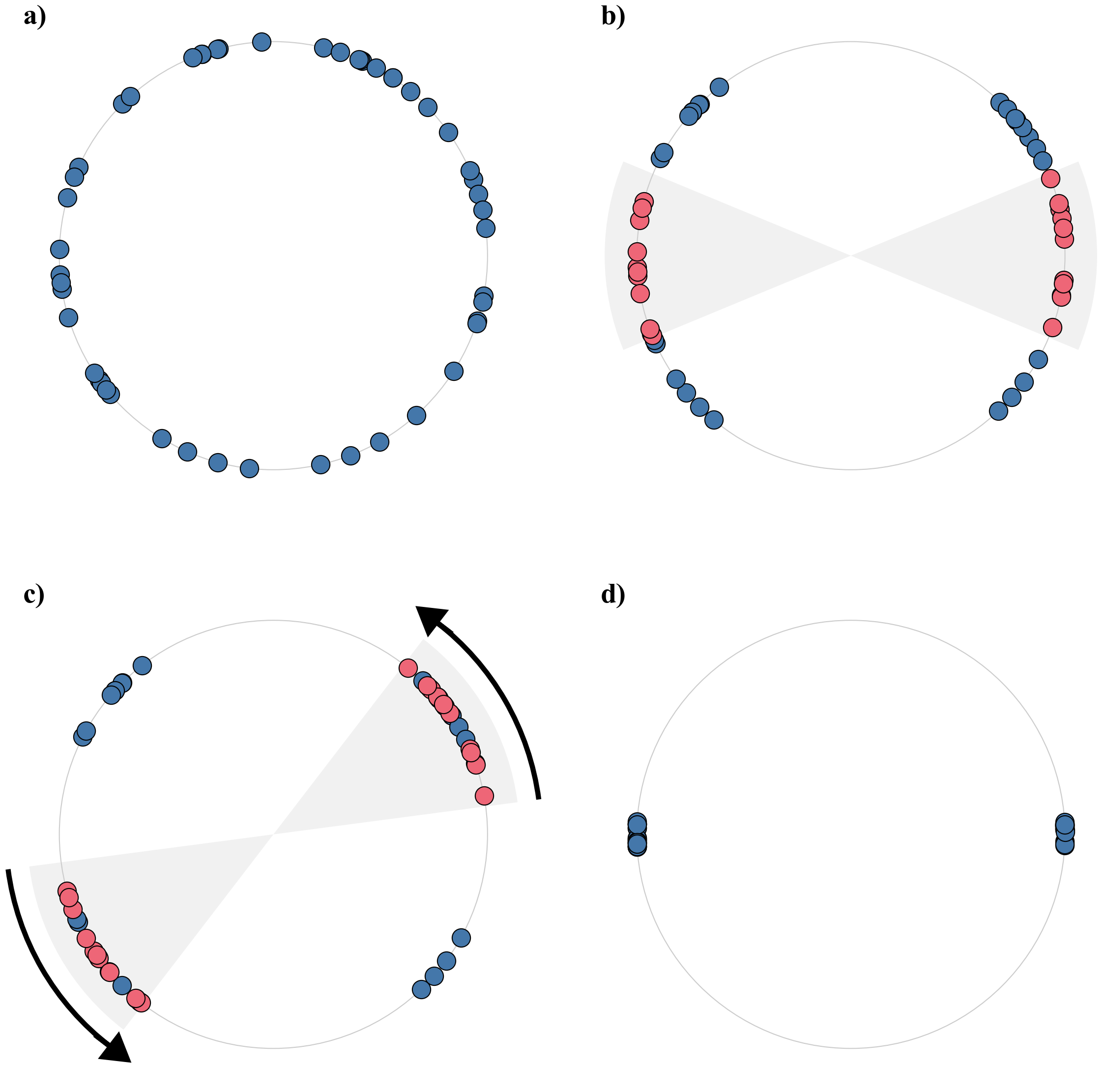}% <-- replace path/name
\caption{Visualization of Cosm dynamics with a dual window twist (DWT) perturbation. Variables are shown as filled circles residing on a unit circle. a) The system is initialized to a uniform random state. b) In addition to variable movements due to sequential conflict-free search, at regular intervals a new pair of diametrically opposed windows (shaded) is positioned according to a random reference angle $\theta_{ref}$. c) Variables within the dual windows rotate by an amount $r_\mathrm{DWT}$ times the current step size $\alpha_t$, after which the dynamics continue. d) Illustrative system configuration after an annealing run with multiple sweeps and DWT perturbations.}
\label{fig:dynamics_with_dwt}
\end{figure}

In the Ising realization studied here, the deterministic SCS rules are complemented by a novel correlated perturbation of the circular variables. The intent is to perturb the system such that sets of variables that have emerged with members at opposite regions of phase space, which we refer to as \emph{polarized clusters}, can be perturbed together and made more likely to separate and rotate freely to lower energy configurations. At occasional times during system evolution, Cosm selects two equal-sized windows at diametrically opposed locations determined by a random reference angle $\theta_{ref}$. All variables within either window rotate by the same magnitude and in the same direction. We refer to this as a \emph{dual window twist} (DWT) perturbation. No information is needed about the existence or location of polarized clusters. In the implementation used here, perturbation timing follows a deterministic schedule in which perturbations occur every $T_\mathrm{DWT}$ sweeps. DWT is a structured correlated perturbation with stochasticity introduced through the random reference angle. The DWT perturbation concept is illustrated in Fig.~\ref{fig:dynamics_with_dwt} and pseudocode is provided in Alg.~\ref{alg:cosm}. DWT perturbations and an example of coordinated cluster motion are visible in Fig.~\ref{fig:varTrajectory1}c,d.

By rotating dual opposing windows rather than a single window, the achieved alignment and anti-alignment relationships among variables within the windows are approximately maintained (Fig.~\ref{fig:dynamics_with_dwt}b,c). The width of the windows is a tunable parameter; we find empirically that a width of 90 degrees (leading to exactly half of the phase space being included in the twist) is effective on the problems tested. The direction of the twist (clockwise or counter-clockwise) can be randomized, though with 90-degree windows and without local bias fields, we find that use of a single direction is sufficient due to symmetry. The DWT step size (magnitude of rotation) is changed over time according to a schedule; in this work, we set the DWT step size to a constant factor $r_\mathrm{DWT}$ of the instantaneous global step size $\alpha_t$. 

In concert with Cosm's other features, the DWT mechanism is meant to promote energy-improving many-variable moves. Achieving useful cluster moves has long been an aim in binary optimization. Among spin-glass simulation methods, replica-based cluster methods such as the Houdayer algorithm~\cite{houdayer2001} and isoenergetic cluster moves~\cite{zhu2015} can greatly accelerate simulations in some regimes. These involve comparing the full spin configurations of multiple replicas, explicitly constructing clusters from the overlap, and flipping the binary states of selected clusters. In Cosm, by contrast, clusters of variables in a continuous circular space can emerge, separate, and move in a single dynamical system.

\subsection{Complete algorithm}
\label{sec:pseudocode}

\begin{algorithm}[t]
\caption{Cosm}
\small
\label{alg:cosm}
\begin{algorithmic}[1]
\State Obtain a proper edge coloring of the input graph
\State Initialize each $x_i$ independently and uniformly at random on $[0,360^\circ)$
\For{$sweep = 1$ to $N_\mathrm{sweeps}$}
  \For{$c = 1,\ldots,C$}
    \For{each edge $(i,j) \in E_c$}
      \State Update $x_i, x_j$ according to equation~(\ref{eq: pair_interaction_contributions})
    \EndFor
  \EndFor
\If{{$sweep$ is a multiple of $T_\mathrm{DWT}$}}
\State Select random reference angle 
$\theta_\mathrm{ref}$
\State Define a pair of diametrically opposed windows using $\theta_\mathrm{ref}$
\For{each $x_i$ within the windows}
\State Rotate $x_i$ by $r_\mathrm{DWT}\alpha_t$ 
\EndFor
\EndIf
\State Update $\alpha_t$ for the next sweep according to the step-size schedule
\EndFor
\For{$p$ = 1 to $N_\mathrm{bisectors}$}
\State Generate a candidate bisector of the unit circle
\State  Binarize $x_i$ into $s_i$ using the two semicircles
\State Evaluate the original binary objective
\State Retain the best objective value and corresponding binary solution $s$
\EndFor
\State Return the best objective value and corresponding binary solution $s$
\end{algorithmic}
\end{algorithm}

The Cosm algorithm for sparse Ising optimization is described in pseudocode in Alg.~\ref{alg:cosm}. It consists of objective compilation using a proper edge coloring, a main search loop of $N_\mathrm{sweeps}$ sweeps, and a final step that converts the continuous variables to binary (i.e., binarization). The coloring may be generated during preprocessing or reused when a precomputed coloring is available. With 2D and 3D lattice graph instances, an edge coloring can be trivially assigned using a regular pattern. With non-lattice graph instances, many polynomial-time edge-coloring heuristics are possible. In this work, we use a simple deterministic greedy heuristic. Edges are processed in a fixed order, with each edge assigned the lowest available color. The edge ordering used in the experiments is specified in Supplementary Section~\ref{sec:exp_details_Gset}. Each sweep incorporates the SCS updates and, according to the perturbation schedule, may include a DWT perturbation. In this work, the global step size $\alpha_t$ is annealed linearly from $\alpha_0$ to zero over the course of a run. After annealing, the unit circle is bisected into two halves, and variables in one half are assigned $-1$ and in the other $+1$. Problems that have local bias fields have defined semicircular sections and thus this binarization step is trivial. For problems that do not have local bias fields, a procedure is needed to find an effective bisector. Here, we employ a set of equally-spaced bisectors and select the best resulting solution (an alternative algorithm that finds the optimal bisector is provided in Ref. \cite{burer2002}).

\begin{figure*}
    \centering
    \includegraphics[width=\textwidth]{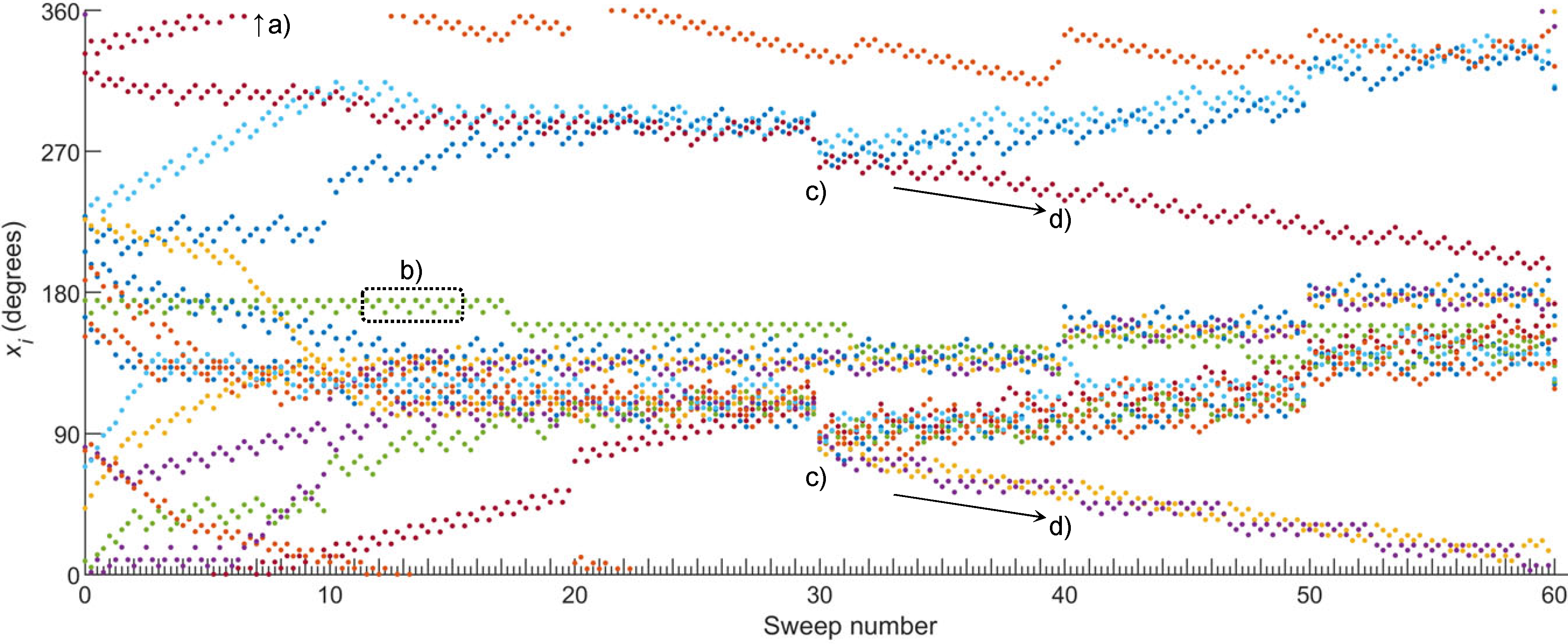}
    \caption{Example Cosm dynamical system evolution. a) Variables wrap around in the circular phase space. b) Example dynamics resulting from sequential conflict-free search (SCS) and sign-based pairwise interactions, including intra-sweep fluctuations. The encircled trajectory spans three sweeps with four subsweeps (edge-color partitions) per sweep. c) DWT perturbations are visible as discontinuities after every 10th sweep, affecting variables within randomly placed dual 90-degree windows separated by 180 degrees. d) Example of a polarized cluster move beginning near sweep 30 when three variables (one in the upper half of phase space and two in the lower) decrease their angular position together through sweep 60; the three variables (3, 6, and 7 using 1-based numbering) represent a contiguous, coupled chain in the problem graph. The example is a 16-variable instance from the planted-solution benchmark set described in Section~\ref{sec:benchmark_problems} (2D toroidal spin glass, L=4, instance no. 31, $J_{ij} \in \{\pm 1, \pm 2\}$~\cite{hou2025}); the first 60 sweeps of a 300-sweep anneal are shown.}
    \label{fig:varTrajectory1}
\end{figure*}

Figure~\ref{fig:varTrajectory1} illustrates how the complete Cosm search process gives rise to collective dynamics during an annealing run.

\subsection{Computational structure}

The compilation of the Ising objective into conflict-free subobjectives provides computational structure in addition to defining the search dynamics. Because the interactions within each subobjective share no variables, they can be evaluated without write conflicts or dependencies among updates. The required objective compilation is also lightweight: for the non-lattice Gset instances studied here, the greedy edge-coloring heuristic described in Section~\ref{sec:pseudocode} completes preprocessing in tens of milliseconds (Section~\ref{sec: exp_protocols}). The Ising realization considered here further admits simple sign-based interaction rules, branchless updates, and 16-bit integer arithmetic. The switching, annealing, and perturbation timing can be predetermined, so the set and ordering of local interactions do not depend on the evolving solution state. Random-number generation is required only for initialization and DWT reference angles, not for individual local interactions. These properties reduce control-flow irregularity and make the inner search loop amenable to efficient realization on conventional processors.

The same structure exposes parallelism at multiple levels. Within a replica, the conflict-free interactions of an active subobjective are mutually independent and can in principle be evaluated concurrently. Across replicas, distinct search trajectories share the same graph structure, subobjective schedule, and annealing schedule, allowing multiple replicas to execute in lockstep and enabling SIMD vectorization across replica state. Additional replicas or SIMD batches can be executed concurrently across processor cores. Thus, although the collective search trajectory can be complex and replica dependent, the operations generating it remain simple and regular.

\section{Results}
\label{sec:results}

We evaluate Cosm on sparse Ising benchmarks (Section~\ref{sec:benchmark_problems}) spanning 2D spin-glass instances with planted solutions and large instances from the longstanding Gset Max-Cut benchmark suite. The Gset experiments include the three largest low-degree random graphs (G60, G61, G70), three of the largest 2D spin glasses (G72, G77, G81), and G64, which has a highly heterogeneous degree distribution and the maximum number of edges across the Gset suite.

\subsection{Ground-state attainment and empirical scaling on 2D spin glasses with planted solutions}

\begin{figure}[b]
\centering
\includegraphics[width=1\columnwidth]{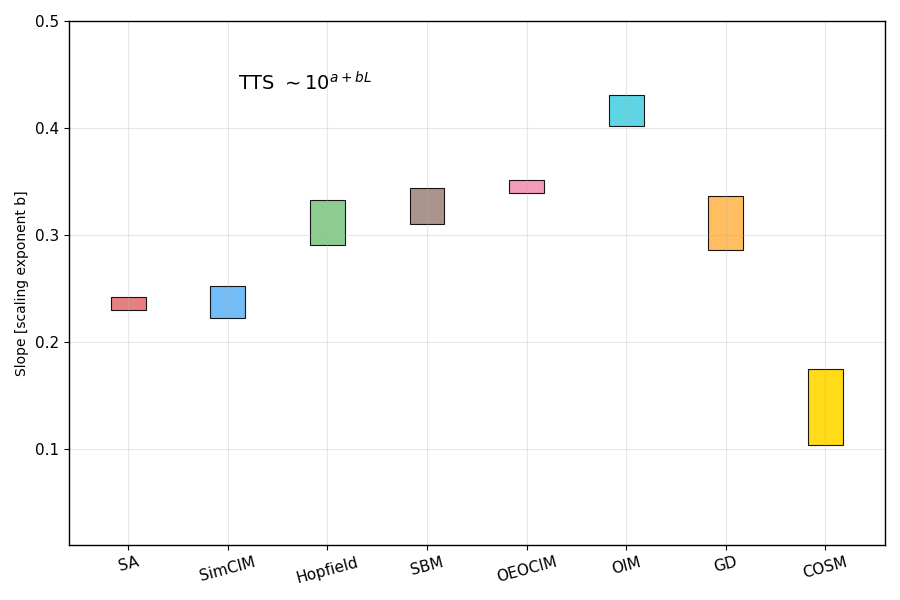}% <-- replace path/name
\caption{Comparison of $TTS$ scaling exponents of eight solvers on planted-solution benchmark instances. Error bars are shown for seven other solvers; data were extracted from Ref. \cite{hou2025preprint} and re-plotted and should be considered approximate. Cosm's 95\% confidence interval is shown on the right. }
\label{fig:chook-boxplot}
\end{figure}

We first evaluate Cosm on the planted-solution 2D spin-glass benchmark previously used by Hou et al. to compare several prominent dynamical solvers~\cite{hou2025}. Hou et al. report that many of these solvers were unable to find the ground states at the larger problem sizes. The comparative scaling results in Figs. 1 and 2 of Hou et al.~\cite{hou2025} are based on data shown through $L=20$. In contrast, Cosm attains the known ground state of every instance across the full available size range, $L=4$--$32$ ($n=16$--$1024$). Specifically, at $n=1024$ with 10,000 trials and $N_\mathrm{sweeps}=1000$, the median $P_\mathrm{s}$ across the 100 instances is $2.09\%$. 

For direct comparison, we follow Hou et al.'s scaling methodology and fit Cosm's $TTS$ over the same five largest problem sizes shown in their Fig. 1 ($L=12,14,16,18,20$)~\cite{hou2025}. The instance with the median $P_\mathrm{s}$ was identified at each size, and execution times were measured using an unoptimized C implementation of Cosm. The corresponding time-to-solution $TTS_{99}$ values are 6.101 ms, 14.589 ms, 28.739 ms, 53.492 ms, and 78.159 ms, respectively. To match the methodology described in Ref.~\cite{hou2025}, we fit $log_{10}TTS$ data to $a+bL$ and obtained an estimated slope of 0.139, representing the scaling exponent $b$. The coefficient of determination $R^2$ is 0.981 and the lower and upper 95\% confidence bounds are 0.104 and 0.175. This 0.139 value is lower than the scaling exponents reported for the seven previously characterized solvers, including simulated annealing and SimCIM (0.22--0.25). The data are plotted in Fig. \ref{fig:chook-boxplot}. Additionally, we characterized Cosm scaling over a wider range of sizes ($L\in\{12,14,\cdots,32\}$), and obtained a scaling exponent of 0.128.

\subsection{Certified optima on large 2D Gset spin glasses}

We next evaluate Cosm on G72, G77, and G81, three large 2D spin-glass instances that include the two largest problems in Gset by number of variables. For each instance, Cosm attains a better solution than previously reported heuristic methods, reaching cuts of 7008, 9940, and 14060, respectively. Cuts of 7008 and 9940 for G72 and G77, respectively, were previously reported by Ikuta et al. using the exact solver CPLEX~\cite{ikuta2015}, while heuristic studies predating Cosm's results attained lower values. The G72 value was certified optimal in Ikuta et al. For G81, the largest instance in Gset (20,000 variables), Ikuta et al. reported a CPLEX lower bound of 14054 from a run with a time limit of approximately 28 hours~\cite{ikuta2015}. Subsequently, Shylo et al. reported reaching a cut of 14056 with a heuristic method~\cite{shylo2017}. A Cosm solution bitstring corresponding to a cut of 14060 is provided in Ref.~\cite{zick2025} to allow independent validation.

We also supplied the three instances to the exact solver Gurobi 12.0.3. The Gurobi runs independently confirm the optimality of the G72 cut of 7008 and certify the G77 and G81 cuts of 9940 and 14060, respectively, as optimal. Recently, independent work has also attained these cuts heuristically and certified their optimality through matching upper bounds~\cite{glass2026}, while Aadit et al. independently attained the G81 optimum~\cite{aadit2026}.  

Across 1024 trials, Cosm attained the certified optimum 405, 237, and 20 times for G72, G77, and G81, respectively. Attaining the certified optima of all three instances through heuristic search is particularly notable given their approximately 25-year history as challenging optimization benchmarks.

\begin{figure}[h]
\centering
\includegraphics[width=0.4\textwidth]{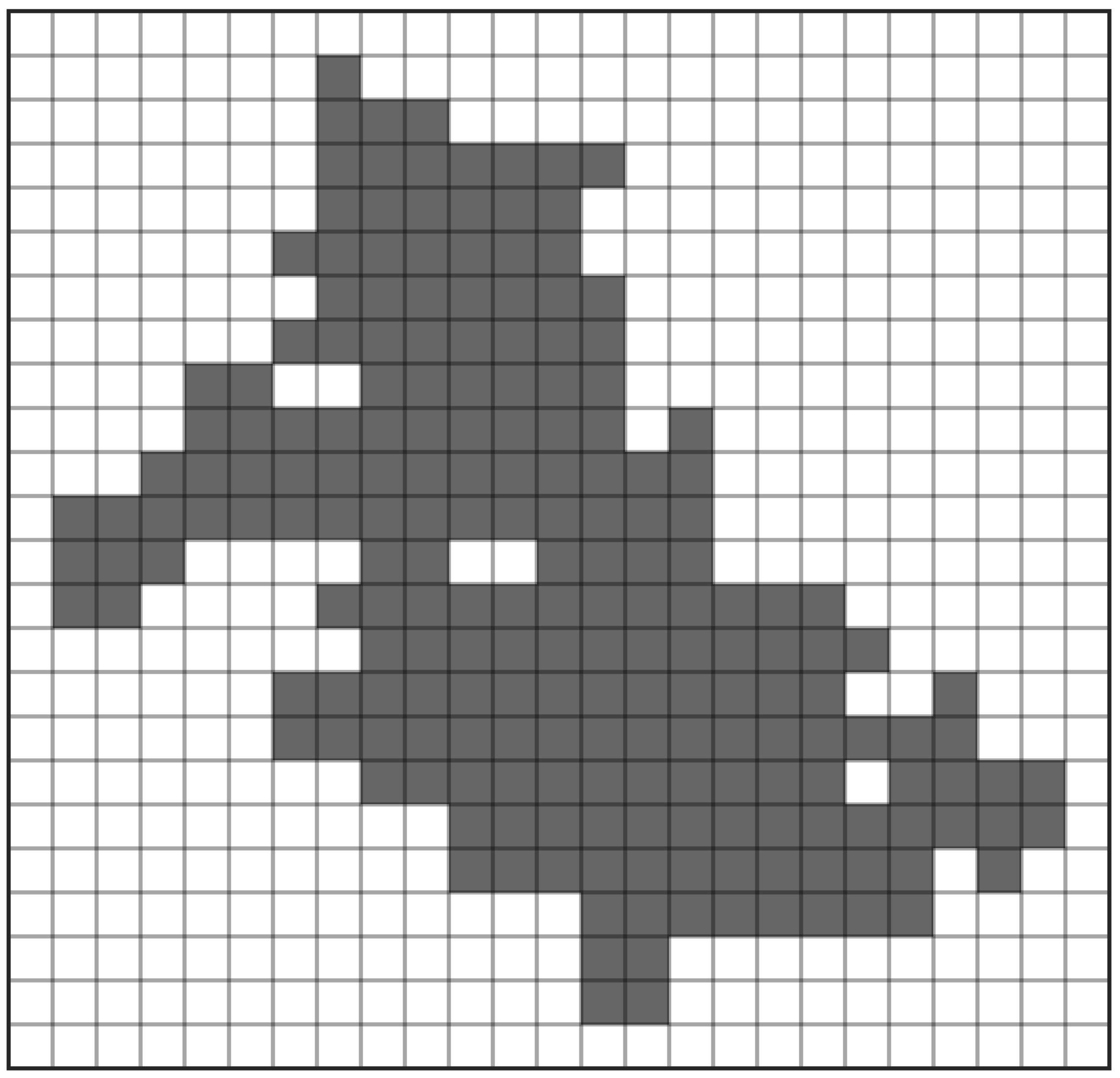}% <-- replace path/name
\caption{Spatial layout of a large cluster of variables (shaded) representing the final move enabling Cosm to reach the optimal solution to Gset instance G72 (7008). Adapted from Ref.~\cite{ZickPatentApp2024}. The cluster is visualized on a 24x25 subset of the 100x100 toroidal problem graph. Each square cell represents a variable. The cluster size is 204 variables, representing over 2\% of all problem variables. The cluster has 128 pairwise bonds at its interface (connections to non-shaded cells). Variables in the cluster each changed their continuous state to reach the opposite binary state, increasing the number of satisfied bonds at the interface from 63 to 65 and improving the cut from 7006 to 7008.}
\label{fig:flippedCluster}
\end{figure}

To gain empirical insight into Cosm dynamics, we previously examined several successful G72 trajectories~\cite{ZickPatentApp2024} in detail. By assigning binary values at intermediate points (normally only performed at the end of an anneal) and testing which of the many variable transitions were consequential, we identified the critical state transition in each trajectory allowing Cosm to exceed a cut of 7006 and reach the certified optimum of 7008. We found that, in some trials, near-simultaneous flips of multiple variable clusters in separate regions of the toroidal graph were responsible. In four trials, a flip of a single contiguous cluster was responsible; the sizes of those clusters were 29, 35, 122, and 204 variables. These observations provide empirical evidence of non-local collective transitions during the search process. Note that a cluster of 204 variables represents more than 2\% of the entire system. The layout of the observed 204-variable cluster on the problem lattice is illustrated in Fig. \ref{fig:flippedCluster}. The cluster has 128 pairwise bonds at its interface (i.e., with variables outside of the cluster). When the 204 continuous variables flipped, the number of satisfied bonds at the interface increased from 63 to 65, improving the cut from 7006 to 7008. These trajectories demonstrate that Cosm can exhibit non-local transitions involving contiguous clusters of variables, suggesting one way in which collective motion may facilitate escape from metastable configurations in large sparse systems.

\subsection{Ablation of Cosm search mechanisms on G70}

\begin{figure}[h]
\centering
\includegraphics[width=0.8\columnwidth]{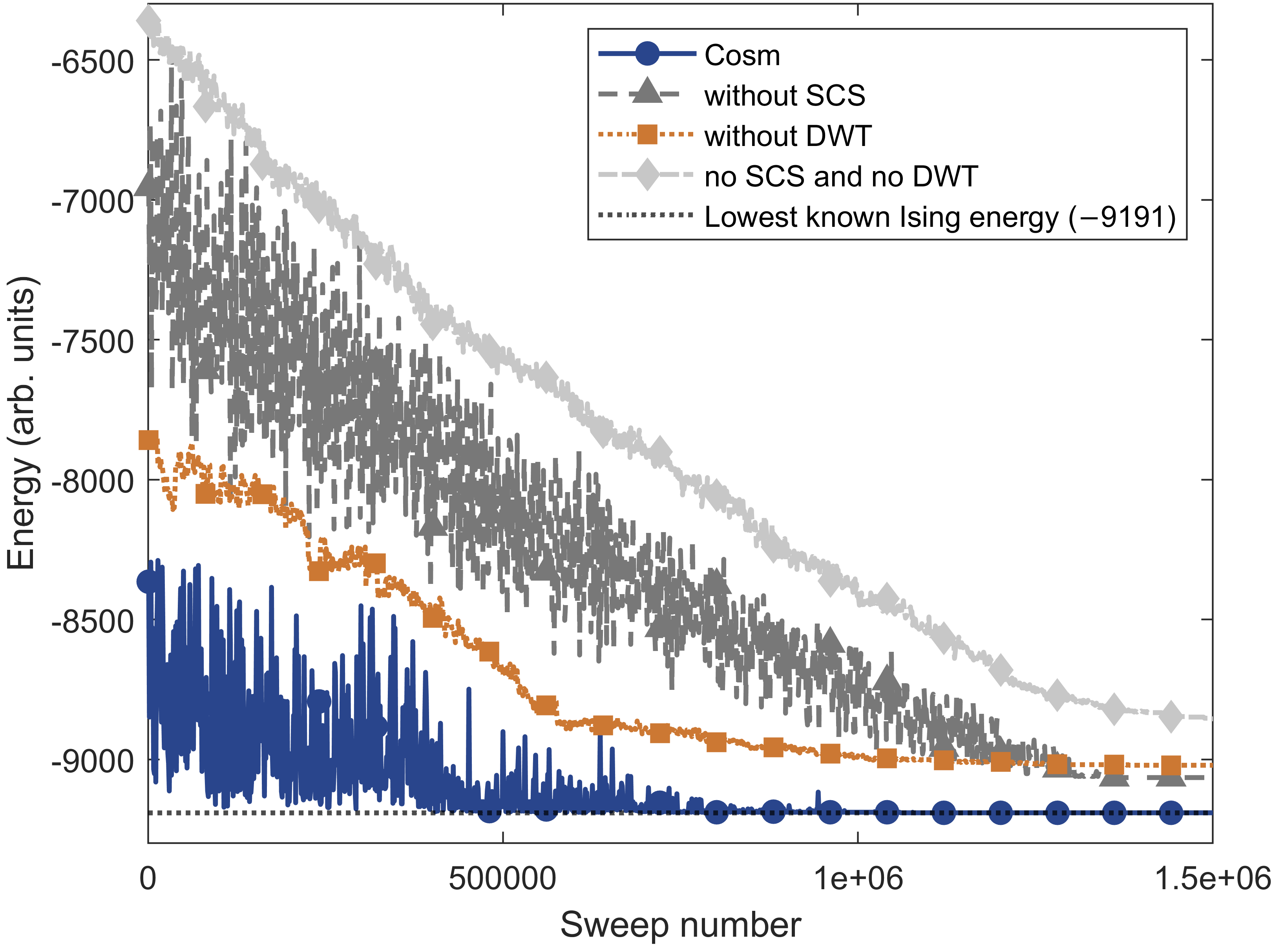}% <-- replace path/name
\caption{Example energy evolution for Cosm and three alternatives tested on instance G70 in ablation study. System energy was sampled once every 1000 sweeps. In this run, Cosm reached the lowest known Ising energy ($-9191$), corresponding to the best-known cut (9595).}
\label{fig:energyTrajectory1}
\end{figure}

To assess the extent to which two of Cosm’s key features contribute to performance, we conducted an ablation study comparing Cosm against three alternative algorithms for which SCS is disabled, DWT is disabled, and both SCS and DWT are disabled, respectively. The results are shown in Table~\ref{tab:ablation}. Across 1000 trials with $N_\mathrm{sweeps}=1.5\times10^6$, Cosm achieves the best-known solution in 16.3\% of trials and an average cut of 9593.39. In the three other cases, the highest achieved cuts are well below the best known, and the average cuts are substantially lower than for Cosm. Example energy trajectories for the four variants are provided in Fig.~\ref{fig:energyTrajectory1}; pronounced energy fluctuations are visible in the two variants employing DWT. Under the tested conditions, removing either SCS or DWT substantially reduces solution quality relative to the full algorithm, while disabling both produces a still larger degradation. These results provide evidence that both mechanisms contribute materially to Cosm's performance on G70.

\begin{table}[htbp]
\centering
\caption{Ablation experiment results. 1000 trials of instance G70. Parameter settings: $N_\mathrm{sweeps}=1.5\times10^6$; $T_\mathrm{DWT}=6$ sweeps. Cosm finds the best-known solution on 16.3\% of trials. Performance drops precipitously when disabling sequential conflict-free search (SCS) or dual window twist (DWT) perturbations.}
\label{tab:ablation}
% Resize implies we force it to textwidth, ensuring it fits perfectly.
%\resizebox{0.8\columnwidth}{!}{%
\begin{tabular}{l c c c}
\toprule
\textbf{Algorithm} & \textbf{Highest cut} & \textbf{Avg cut} & \textbf{\textbf{$\boldsymbol{P_{\text{s}}}$}} \\
\midrule
Cosm & 9595 & 9593.39 & 0.163 \\
without SCS & 9551 & 9528.00 & 0 \\
without DWT & 9544 & 9501.40 & 0 \\
no SCS, no DWT & 9481 & 9432.81 & 0 \\
\bottomrule
\end{tabular}
\end{table}

\subsection{Best-known solutions and time-to-target on low-degree random graphs}
\label{sec:randomGset}

We next consider the large low-degree random-graph instances G60, G61, and G70, for which prior heuristics have had difficulty attaining the best-known solutions. Cosm reaches the best-known solutions for each of the three tested instances---14190 for G60, 5799 for G61, and 9595 for G70. A comparison of Cosm solution quality with other notable results is given in Supplementary Table~\ref{tab:cut_comparisons}. With sufficiently long runs, Cosm attains these challenging targets with substantial success probabilities, rather than only as rare outcomes. For instance, with $N_\mathrm{sweeps}=128\times10^6$ sweeps, the best-known solutions are attained in 27.1\%, 25.7\%, and 70.6\% of trials for G60, G61, and G70, respectively (Table~\ref{tab:success_probability}).

\begin{table}[htbp]
\centering
\caption{Cosm success probability with extended run lengths ($N_\mathrm{sweeps} = 128 \times 10^6$).}
\label{tab:success_probability}
% Resize implies we force it to textwidth, ensuring it fits perfectly.
%\resizebox{\textwidth}{!}{%
%{
\begin{tabular}{c c c}
\toprule
\makecell{\textbf{Gset}\\ \textbf{instance}} &
  \textbf{Successes/trials} &
  \textbf{$\boldsymbol{P_{\text{s}}}$} \\
\midrule
G60 & 277/1024 & 0.271 \\
G61 & 263/1024 & 0.257 \\
G70 & 723/1024 & 0.706 \\
\bottomrule
\end{tabular}%
\end{table}

Before characterizing time-to-target ($TTT_{99}$), $r_\mathrm{DWT}$ was tuned to 1.5 for G60/61, and 1.0 for G70. $N_\mathrm{sweeps}$ was tuned per instance to achieve a success probability of at least 99\% from a single 1024-trial wave. Apart from $r_\mathrm{DWT}$ and $N_\mathrm{sweeps}$, all other solver parameters were held fixed. $\alpha_0$ was set to 14 degrees for each instance.

We find that the Cosm implementation achieves $TTT_{99}$ values measured in seconds for each instance, as shown in Table~\ref{tab:nontorus_TTT}.

Using the Fujitsu second-generation Digital Annealing Unit (DAU), Matsubara et al. reported 99\%-probability times-to-solution of 9831 s and 893 s for attaining the same G60 and G61 targets, respectively~\cite{matsubara2020,takemoto2021}. The corresponding Cosm $TTT_{99}$ values of 9.46 s and 15.05 s are approximately 1040-fold and 59-fold lower, respectively. The DAU natively supports up to 8192 variables; Matsubara et al. did not report a result for the 10,000-variable G70 instance. MOH reported a G60 cut of 14190 with an associated computation time of 2945 s~\cite{ma2017}, although this is not a 99\%-probability time-to-target measurement. Differences in hardware and experimental protocols preclude interpreting these cross-study comparisons as controlled speedup measurements.

On G70, Chen et al. reported attaining a cut of 9595 using a Monte Carlo gradient-based heuristic on an NVIDIA A100 GPU, following prolonged execution for which the time to attainment was not reported~\cite{chen2023preprint}. During standard testing, it reached a cut of 9572~\cite{chen2025}. Goudet et al. reported an evolutionary algorithm reaching 9595 during extended calibration runs, although the target was not reached in ten standard test runs with a 20-hour time limit~\cite{goudet2024}. The Simulated Bifurcation Machine (SBM) attained a wall clock $TTT_\mathrm{99}$ of 31{,}599~s to reach a cut value of 9578 on G70 using an NVIDIA V100 GPU~\cite{goto2021}. Among the prior studies identified, none reported a time-to-target measurement for the G70 cut of 9595. Cosm reaches the best-known G70 cut of 9595 with $TTT_{99}=2.43$ s. 

In addition to $TTT_{99}$, we characterize energy-to-target using measured CPU and DRAM power consumption. Results are provided in Supplementary Section~\ref{sec:ETT}. For example, regarding the G70 best-known solution, the Cosm implementation demonstrates a CPU energy-to-target of 1.24 kJ under the tested configuration.

\subsection{Search performance on heterogeneous-degree G64}
\label{sec:skewGset}

Gset instance G64 provides a substantially different graph structure, with a highly heterogeneous degree distribution (avg. degree 11.8, maximum degree 589). It also has the highest edge count (41,459) in the Gset suite. We find that Cosm reproducibly attains a G64 cut of 8753, which is a new best-known solution. An example Cosm solution bitstring corresponding to a cut of 8753 is provided in Supplementary Section~\ref{sec:G64_8753_bitstring} to allow the cut value to be independently verified. Before the discovery reported here, the best reported G64 cut was 8752~\cite{goudet2024}. Selected prior results are provided in Supplementary Section~\ref{sec: prior_highest_cuts} for comparison.

\begin{figure}[h]
\centering
\includegraphics[width=1.0\columnwidth]{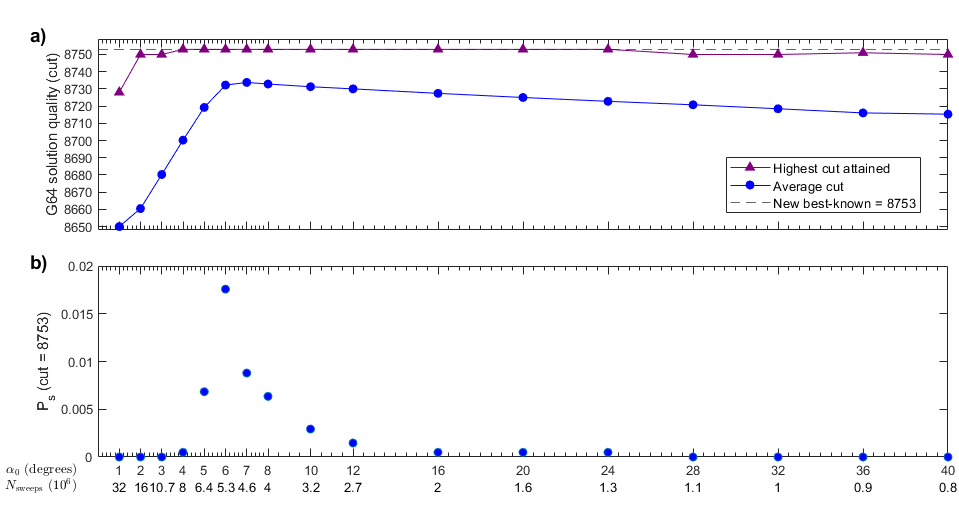}% <-- replace path/name
\caption{Cosm performance on G64 as a function of initial step size $\alpha_0$, with $\alpha_0 N_{\rm sweeps}$ held constant at $32\times10^6$. The corresponding $N_{\rm sweeps}$ values are shown beneath the $\alpha_0$ values. Each point represents 2048 trials. (a) Highest and average cut. (b) Success probability for cut = 8753.}
\label{fig:G64_2panel}
\end{figure}

We next examine how Cosm search efficacy depends on the initial step size $\alpha_0$. To examine the effect of step size while approximately maintaining a common overall update budget, we varied $\alpha_0$ and $N_{\rm sweeps}$ inversely, holding $\alpha_0 N_{\rm sweeps}$ constant at $32\times10^6$. 2048 trials were conducted at each combination of $\alpha_0$ and $N_{\rm sweeps}$. As shown in Fig.~\ref{fig:G64_2panel}a, the new best-known cut is reached over a range from 4 to 24 degrees. The average cut peaks around 6--8 degrees, and drops off sharply below 6 degrees. The success probability peaks around 6 degrees as shown in Fig.~\ref{fig:G64_2panel}b.

To characterize the reproducibility of cut = 8753, parameters were set to $\alpha_0=7$ degrees, $r_\mathrm{DWT}=3.5$, and $N_{\rm sweeps}=10^6$. The parameter selection cost is not characterized here, thus the $TTT$ should be interpreted as the post-tuning time required to reproduce the target, rather than as an end-to-end time for solving a previously unseen instance. Over 128,000 trials, Cosm reaches a $P_\mathrm{s,trial}$ of 0.193\%. 
Measured CPU and DRAM energy-to-target for this configuration are reported in Supplementary Section~\ref{sec:ETT}. The measured success probability and the average wave time of 88.36 s yield a $TTT_{99}$ of 205.8 s (Table~\ref{tab:nontorus_TTT}) for reproducing the newly established G64 best-known cut of 8753.

\begin{table}[htbp]
\centering
\caption{Cosm performance on large random-graph and heterogeneous-degree Gset instances.}
\label{tab:nontorus_TTT}
\begin{tabular}{c r l c c r r}
\toprule
  \makecell{\textbf{Gset}\\ \textbf{instance}} & 
  \makecell{\textbf{Target}\\ \textbf{solution}} & 
  \textbf{$\boldsymbol{P_\mathrm{s,trial}}$} &
  \makecell{\textbf{Calc.}\\ $\boldsymbol{P_\mathrm{s,wave}}$} &
  \textbf{$\boldsymbol{R_\mathrm{99,wave}}$} &
  \makecell{$\boldsymbol{t_\mathrm{wave}}$\\ \textbf{(s)}} &
  \makecell{$\boldsymbol{TTT_{99}}$\\ \textbf{(s)}} \\
\midrule
G60 & 14190 & 651/128000=0.00509 & 0.9946 & 1.0 & 9.46$\pm$0.38 & 9.46 \\
G61 & 5799 & 668/128000=0.00522 & 0.9953 & 1.0 & 15.05$\pm$0.38 & 15.05 \\
G64 & 8753 & 247/128000=0.00193 & 0.8616 & 2.329 & 88.36$\pm$1.49 & 205.8 \\
G70 & 9595 & 611/128000=0.00477 & 0.9926 & 1.0 & 2.43$\pm$0.31 & 2.43 \\
\bottomrule
\end{tabular}
\end{table}

\section{Relationship to existing approaches}

Cosm has conceptual connections to several established approaches, including dynamical Ising solvers, incremental optimization methods, and conflict-free scheduling.

Physics-inspired optimization approaches such as SBM and OIM use interacting dynamical variables to search for low-energy configurations~\cite{goto2021,sreedhara2024}. Digital memcomputing similarly exploits collective behavior arising from engineered nonlinear dynamics~\cite{diventra2019}. Cosm shares this emphasis on dynamical computation while treating the temporal organization of interactions as a central design principle. While many oscillator-based formulations employ smooth coupling functions, the present Cosm realization uses sign-based interactions whose magnitudes are independent of angular separation. Sign-based coupling is also employed in the dSB variant of SBM~\cite{goto2021} and a recent OIM formulation~\cite{sreedhara2024}.

Incremental and stochastic-gradient methods share with Cosm the use of updates derived from a decomposed objective. These methods may apply updates from individual components or minibatches, using cyclic or stochastic selection. Sequential component updates can produce trajectories that differ from aggregate-gradient updates, particularly at larger step sizes. In Cosm, this temporal composition is deliberately exploited through conflict-minimized subobjectives and combined with structured correlated perturbations to promote collective search behavior. Block-coordinate methods update subsets of variables while retaining the aggregate objective, whereas Cosm decomposes the objective into subsets of interactions. Parallel coordinate-descent methods have used edge coloring to schedule matching-based sets of nonconflicting pairwise updates~\cite{choi2022efficient}. Vertex-colored update schemes in probabilistic-bit methods provide a separate example of conflict-free computation~\cite{chowdhury2025}.

Dynamic-neighborhood cellular automata~\cite{dantchev2011} and synchronization on time-varying networks~\cite{belykh2004,wang2025} provide related examples of time-varying interaction structures. Guided self-organization seeks to facilitate useful macroscopic emergent behavior through appropriate forms of guidance or control~\cite{gershenson2020}.

\section{Discussion}

\subsection{Empirical interpretation}
The results indicate that Cosm's strong search behavior is not confined to a single benchmark family or graph structure. Scaling results on spin-glass instances with planted solutions are accompanied by attainment of certified optima on large two-dimensional Gset problems, reliable attainment of best-known solutions on large random-graph instances, and a new best-known solution on the heterogeneous-degree G64 instance. The breadth of these results provides empirical evidence that the Cosm search framework can be highly effective across substantially different sparse Ising landscapes. The G64 result further shows that this behavior is not restricted to low-degree structures.

\subsection{Mechanistic interpretation of SCS and DWT}
The non-local cluster transitions observed in the G72 trajectories (Fig.~\ref{fig:flippedCluster}) provide direct empirical evidence that collective motion can arise during Cosm search. 

SCS does more than separate conflicting interactions: sequential activation of the resulting subobjectives creates dynamics that appear important to its search behavior. Because each subobjective acts on a state modified by preceding subobjectives, a sweep generally produces a trajectory different from one generated by the aggregate influence of the same interactions. In the G64 experiments, search efficacy depends strongly on the initial step size $\alpha_0$. Reducing $\alpha_0$ while proportionally increasing $N_\mathrm{sweeps}$ did not preserve performance, and the best solution quality was obtained at an intermediate step size. These observations suggest that the global update scale does more than control the magnitude of individual updates; it may influence the dynamical regime in which search occurs. Effective search may require an intermediate range of step sizes, large enough for temporal composition to have substantial effects without becoming excessively disruptive. Understanding how finite-step temporal composition shapes Cosm's collective dynamics, and how these dynamics relate to aggregate updates in the small-step limit, remains an open theoretical challenge. 

DWT complements SCS by using a global cue to induce correlated, state-dependent variable movements (Fig.~\ref{fig:varTrajectory1}c,d). Scaling the DWT perturbation magnitude with the global step size, as we have done here, may help maintain a useful balance between local interaction updates and correlated perturbations throughout annealing. The G70 ablation results show that, under the tested conditions, the best-known solution was not reached without DWT.

The instances tested here have discrete weights, which can increase the prevalence of exact cancellation among competing interactions, leading to zero-local-field spins and potentially greater degeneracy. This distinction may matter for collective optimization methods: recent results for the cluster-based message-passing (CluMP) optimizer indicate that degeneracy can reduce the effectiveness of collective updates relative to continuously weighted spin glasses~\cite{rissone2026clump}. Cosm may respond differently because SCS temporally separates competing local influences. This raises the possibility that switched dynamics may be relatively effective on landscapes characterized by frequent local-field cancellation or degeneracy. Because competing interactions can also nearly cancel with continuous weights, any benefits of SCS need not be restricted to discrete-weight or exactly degenerate landscapes. Controlled comparisons across weight distributions and the extent of local-field cancellation are needed to investigate this hypothesis.

\subsection{Applicability}
The strong results across diverse sparse interaction structures motivate investigation of application domains in which sparsity arises from discrete constraints. Error-correcting-code decoding is one candidate: LDPC and related codes are defined by sparse graphs of parity constraints, and decoding places a premium on rapid search among discrete configurations. Other candidates include graph partitioning, scheduling, and constraint-satisfaction problems that admit sparse Ising or QUBO formulations. Such applications may introduce higher-order constraints, auxiliary variables, heterogeneous coupling strengths, and graph structures not represented by the benchmarks considered here. The circular-variable formulation may also be relevant to training neuromorphic systems with phase-valued parameters, such as metasurface optical classifiers~\cite{leonard2021}. Determining whether Cosm's advantages transfer to these application domains—and which problem characteristics determine its effectiveness—remains an important direction for future work.

\subsection{Limitations}
The present results do not yet establish the problem characteristics that determine when Cosm provides its strongest advantage. The roles of connectivity, degree heterogeneity, frustration, coupling distribution, degeneracy, and other landscape characteristics remain to be determined. The present study also demonstrates only an Ising realization of the general Cosm framework, and its effectiveness for other optimization formulations remains to be established. Within the Ising setting, the experiments are limited primarily to synthetic benchmarks; further work is needed to determine whether the observed advantages transfer to application-derived problems and broader coupling distributions. 

The reported times-to-target characterize execution after parameter selection and therefore do not include offline tuning cost. Common values of the principal dynamical parameters were effective across several structurally related instances, and the selected DWT settings exhibit a suggestive pattern: $r_{\rm DWT}=0.5$, 1.0, and 1.5 were used for Gset groups with maximum degrees 4, 9, and 14, respectively, while the highly degree-heterogeneous G64 instance used a larger value of 3.5. Although these limited observations do not establish a predictive relationship, they raise the possibility that useful parameter settings could be inferred from problem characteristics. An important open question is how well Cosm performs end-to-end on previously unseen instances, including the computational cost of parameter selection. 

The ablation results establish the importance of SCS and DWT under the tested conditions, but the contribution of observed coordinated cluster transitions to Cosm's search performance has not yet been quantified.

\subsection{Computational architecture}
Cosm combines complex collective search behavior with highly regular computational execution. In the vectorized CPU implementation, replicas following different search trajectories nevertheless share the same graph structure, interaction sequence, and annealing schedule, with local interactions implemented using branchless integer operations. Although the sparse graph produces irregular memory accesses across variables, the access pattern remains aligned across SIMD replicas, enabling lockstep SIMD execution. Thus, different search trajectories can be generated while maintaining uniform control flow and aligned memory access. The compilation step that creates the subobjective schedule is inexpensive; even for G64, greedy edge coloring, reordering, and associated file I/O require only 40 ms. The G60 and G61 results further demonstrate that Cosm can attain challenging solution targets on conventional CPUs with substantially shorter reported times than a prior dedicated optimization processor~\cite{matsubara2020}, although the cross-study comparison does not isolate hardware and algorithmic contributions to the performance differences.

This computational structure may also be well suited to GPUs, with parallelism available both across lockstep replicas and within each replica through the conflict-free interactions of an active subobjective. The sequential dependence among subobjectives requires synchronization, but this overhead can be partially mitigated when sufficient parallel work is available within each subsweep.

\subsection{Broader computational implications}
Cosm illustrates how simple interacting elements can give rise to useful system-level computation through suitably orchestrated collective dynamics. The present realization uses temporal orchestration of subobjectives together with structured correlated perturbations, but these mechanisms represent only a small part of a much larger dynamical design space. Collective dynamical systems can be explored as discovery spaces for additional computational mechanisms that can then be abstracted, combined, and implemented on suitable conventional or unconventional computing substrates. Cosm therefore suggests another way to approach computational search. Finding superior solutions need not begin with a prescribed set of search moves; one can instead design or discover local dynamics and orchestration mechanisms from which effective search emerges.

\section{Conclusion}

We introduced Cosm, a computational framework based on orchestrated collective dynamics, and developed and tested a heuristic realization for sparse Ising optimization. Across sparse Ising benchmarks with varied graph structures, Cosm attains certified optima and best-known solutions, including a new best-known solution on the heterogeneous-degree G64 instance. The planted-solution results suggest favorable empirical scaling relative to previously characterized solvers, while the vectorized CPU implementation demonstrates that these collective search dynamics can be realized efficiently on conventional parallel hardware. The findings suggest promise for further exploring ways to harness collective dynamics for computation.

Cosm demonstrates that, from many simple interacting elements and a new way of orchestrating their dynamics, powerful computational search can indeed emerge. 

\section{Methods}
\label{sec:methods}
\subsection{Benchmark problems}
\label{sec:benchmark_problems}

We first consider a set of tile-planted 2D Ising lattice instances that were previously used by Hou et al. to benchmark several prominent solvers including simulated annealing, oscillator Ising machine, Simulated Bifurcation Machine, and Coherent Ising Machine~\cite{hou2025}. The instances were generated using the Chook benchmark instance generator~\cite{perera2020}, with interaction weights from the set $\{\pm 1,\pm 2\}$; the use of two weight magnitudes allows tuning of the problem hardness~\cite{perera2020hardness}. The problem graphs have degree-4 nearest neighbor connectivity and periodic boundaries in both dimensions. By construction, using the tile-based approach, each instance has a known ground state making the set suitable for benchmarking. The collection consists of 100 instances for each problem size ranging from 16 to 1024 variables, where the number of variables $N=L^2$ and $L\in\{4,6,8,\dots,32\}$. The problem
instances and ground state energies were provided by one of the authors (H.K.) of Ref.~\cite{hou2025}.

For all other experiments, we use large problem instances from the Gset benchmark suite \cite{gset} which is a canonical collection of sparse Max-Cut problems. Originally described between 1999~\cite{benson1999} and 2000~\cite{choi2000}, it has since been widely adopted to assess the performance of both conventional solvers and physics-inspired Ising machine models. This suite contains graphs of different types, such as square grid with periodic boundaries in both dimensions (2D torus), graphs with low degree and random structure, and skew-random graphs with highly heterogeneous degree distributions. For 2D torus problems, we select the three largest such problems in the Gset suite, which are G72, G77, and G81. These are equivalent to 2D Ising spin glasses with degree-4 nearest neighbor connectivity and bimodal ($\pm J$) couplings. Searching for the ground state of an Ising spin glass defined on a lattice graph (Edwards-Anderson spin glass) is a canonical problem in condensed matter physics \cite{dahlberg2025} as well as an established benchmark for binary quadratic optimization \cite{katzgraber2018}. While many nonplanar Ising problems are NP-hard, the zero-field $\pm J$ Ising model on a square lattice with periodic boundary conditions is polynomial-time solvable~\cite{barahona1982}, and specialized heuristic methods have also been developed for this problem class~\cite{khoshbakht2018}. The three instances considered here contain $10{,}000$–$20{,}000$ variables and have nevertheless remained longstanding heuristic benchmarks. Beyond these lattice-structured problems, we test the three largest instances from Gset's low-degree random-graph category: G60, G61 and G70 (maximum degree of 14, 14, and 9, respectively). These may be considered instances of bounded degree problems for which the coupling degree does not grow with the number of variables, for instance due to physical constraints (e.g., in road networks or VLSI digital circuits). G60 and G61 have identical graph structure but with different weight distributions (uniform vs. random $\pm$1). Finally, we select an instance from Gset's skew-random graph category. We use instance G64 which has a maximum degree of 589, average degree of 11.8, and 41,459 edges which is the largest edge count in Gset.

Empirically, heuristic optimizers have had difficulty attaining the best-known solutions on the low-degree and skew-random benchmark instances. Because difficult targets may require many trials for probabilistic solvers, these instances are more informative than the 2D torus instances for characterizing replica-parallel time-to-target performance. 

The Gset problems can be formulated as Ising problems or, equivalently, as Max-Cut problems
\begin{equation}
\label{eq:maxcut}
\max \frac{1}{2} \sum_{1 \leq i < j \leq n} w_{ij}(1 - x_i x_j)
\end{equation}
with $n$ binary variables $x_i \in \{-1, +1\}$ \cite{burer2002}. When converting between Ising and Max-Cut formulations, $w_{ij} = -J_{ij} $. Note that the Gset problem files use the Max-Cut sign convention for the weights. Key properties of the tested Gset instances are summarized in Table~\ref{tab:problem_characteristics}.

\begin{table}[htbp]
\centering
\caption{Characteristics of the tested Gset benchmark instances. $n$: number of variables, $E_G$: set of edges.}
\label{tab:problem_characteristics}
% Resize implies we force it to textwidth, ensuring it fits perfectly.
%\resizebox{\textwidth}{!}{%
\begin{tabular}{c r r r r l l}
\toprule
\makecell{\textbf{Gset}\\ \textbf{Instance}} &
  \multicolumn{1}{c}{$\boldsymbol{n}$} & % Multicolumn used to center the header over right-aligned data
  \multicolumn{1}{c}{$|\boldsymbol{E_G}|$} &
  \makecell{\textbf{Max}\\ \textbf{degree}} &
  \makecell{\textbf{Avg}\\ \textbf{degree}} &
  \textbf{Weights} & \textbf{Structure} \\
\midrule
G60 & 7000 & 17148 & 14 & 4.9 & Unweighted & Random \\
G61 & 7000 & 17148 & 14 & 4.9 & $\{-1,+1\}$ & Random \\
G64 & 7000 & 41459 & 589 & 11.8 & $\{-1,+1\}$ & Skew \\
G70 & 10000 & 9999 & 9 & 2.0 & Unweighted & Random \\
G72 & 10000 & 20000 & 4 & 4.0 & $\{-1,+1\}$ & 2D torus \\
G77 & 14000 & 28000 & 4 & 4.0 & $\{-1,+1\}$ & 2D torus \\
G81 & 20000 & 40000 & 4 & 4.0 & $\{-1,+1\}$ & 2D torus \\
\bottomrule
\end{tabular}%
\end{table}

\subsection{Performance metrics}
\label{sec:performance_metrics}

Success probability per trial $P_\mathrm{s,trial}$ is the fraction of trials reaching a specified target solution quality. The optimized C implementation of Cosm uses 32 SIMD replicas per job, and is generally run with 32 concurrent jobs on the hardware. We refer to this set of 1024 concurrent replicas as a \emph{wave}. Using the independence approximation validated empirically in Supplementary Section~\ref{sec:exp_details_Gset}, we calculate the success probability of each wave as
\begin{equation}
\label{eq: P_s,wave}
P_\mathrm{s,wave} = 1 - (1-P_\mathrm{s,trial})^{1024}.
\end{equation}

The effective number of repetitions required to reach a target with at least 99\% probability is taken to be
\begin{equation}
R_\mathrm{99,wave}(target) = \begin{cases}1.0, & \text{if } P_\mathrm{s,wave} > 0.99, \\
\frac{\log(1-0.99)}{\log(1-P_\mathrm{s,wave})}, & \text{if } 0 < P_\mathrm{s,wave} \leq 0.99
\end{cases}
\end{equation}
which has a minimum value of 1.0 to stipulate that at least one wave is required. The repetition count is treated as a continuous quantity following the conventional time-to-target formulation; an actual execution using complete waves would require rounding up to the next integer. The time-to-target is defined as the measured average wave time $t_\mathrm{wave}$ times the number of effective repetitions
\begin{equation}
TTT_{99}(target) = t_\mathrm{wave} \cdot R_\mathrm{99,wave}(target).
\end{equation}
In the special case where the target is the optimal solution, we refer to the corresponding metric as time-to-solution ($TTS$). 

\subsection{Experimental protocols}
\label{sec: exp_protocols}
\textbf{Scaling experiment.} We first benchmark Cosm on the aforementioned planted-solution instances and characterize $TTS$ scaling over the same range of problem sizes considered by Hou et al.~\cite{hou2025}, using the same exponential scaling form. Though absolute runtimes are not directly comparable to the prior study due to different codebases and hardware, scaling exponent comparisons are more robust to implementation differences. Annealing times are measured using an unoptimized C implementation of Cosm running on a 2$\times$ Xeon Gold 6544Y CPU with 32 physical cores (2$\times$ 16), 2 TB of memory, and a 3600 MHz clock frequency. For direct comparison, $TTS$ scaling is characterized over the five largest problem sizes shown in Fig. 1 of Hou et al. At each problem size, $N_{\rm sweeps}$ was tuned while the other Cosm parameters were held fixed across sizes. We additionally test all available problem sizes. Experimental details are provided in Supplementary Section~\ref{sec: scaling exp details}. 

\textbf{Preprocessing.} For Ising problems, Cosm objective compilation consists of assigning the input graph an edge coloring and organizing edges by color. For the lattice graph instances, a simple regular 4-color edge coloring is employed. For non-lattice Gset instances, the deterministic greedy edge-coloring heuristic from Section~\ref{sec:pseudocode} is used; the average preprocessing times using a C implementation, including file I/O, were measured across ten samples as $19\pm8$ ms for G60, $19\pm9$ ms for G61, $40\pm15$ ms for G64, and $16\pm8$ ms for G70. Because this per-instance compilation cost is small relative to the measured anneal times, it is excluded from $TTT$. The resulting edge coloring and subobjective schedule can also be shared across replicas and reused across search trials, allowing the cost to be amortized over the solution effort for a single instance.

\textbf{\emph{TTT} experiments.} For $TTT$ experiments on Gset instances, we employ an optimized C implementation of Cosm. The relaxed spin variables are represented as 16-bit unsigned integers, and the inner interaction loop uses branchless integer operations with modular arithmetic exploiting the natural wraparound of the integer representation. The implementation exploits replica-level parallelism by representing the states of 32 replicas within AVX-512 vectors and updating them in lockstep. The 32 replicas maintain separate variable states but share the graph structure, edge-coloring schedule, annealing schedule, and the DWT reference angle generated at each perturbation event. For each benchmark instance, the single-replica success probability $P_\mathrm{s,trial}$ is estimated from 128,000 trials. All wall-clock timing measurements are performed separately on the fully-loaded 32-core $2\times$ Xeon Gold CPU. For each problem instance, 200 wave times are measured. Additional experimental details and solver parameter settings are provided in Supplementary Section~\ref{sec:exp_details_Gset}.

\textbf{Parameter tuning.} For the planted-solution instances, we tuned $N_\mathrm{sweeps}$ for each problem size, and used the same value for all 100 instances at each size. For Gset, $N_\mathrm{sweeps}$ was tuned for each instance. The initial step size $\alpha_0$ and DWT ratio $r_\mathrm{DWT}$ were tuned once per group of structurally related instances, with a common parameter setting used within each group: all planted-solution instances, the G60/G61 pair, and the three Gset 2D torus instances.

\backmatter

\bmhead{Acknowledgements}
The authors thank David Ferguson, Joshua Job, Mohammad Sakib, Itay Hen, Ryan Epstein, Bryan Jacobs, Jonathan Machta, Sergey Novikov, Aaron Pesetski, Marc Sherwin, Federico Spedalieri, Matthew French, Satrant Bains, and Andy Rittenbach for helpful discussions and suggestions. The authors thank Helmut Katzgraber for discussions and for providing the tile-planted problem instances. The authors thank Northrop Grumman Systems Corporation and the University of Southern California's Information Sciences Institute for support. 

\bmhead{Funding}
Nikhil Shukla's contributions were supported in part by National Science Foundation grant \#243387.

\section*{Data Availability}
The benchmark results generated in this study are available from the corresponding author upon reasonable request. Benchmark problem instances are available from the sources cited in the manuscript.

\bibliography{Cosm_references}% common bib file

\bmhead{Author contributions}
K.Z. conceived the study, developed the methodology, implemented the software, conducted the investigation, supervised the research, validated the results, prepared the visualizations, and wrote the original manuscript. N.S. contributed to the formal analysis and reviewed and edited the manuscript. A.M. contributed to the investigation, software implementation, validation, visualization, and review and editing of the manuscript.

\clearpage

% Supplementary numbering
\setcounter{page}{1}
\renewcommand{\thepage}{S\arabic{page}}
\setcounter{section}{0}
\setcounter{figure}{0}
\setcounter{table}{0}
\setcounter{equation}{0}
\renewcommand{\thesection}{S\arabic{section}}
\renewcommand{\thefigure}{S\arabic{figure}}
\renewcommand{\tablename}{Supplementary Table}
\renewcommand{\thetable}{S\arabic{table}}
\renewcommand{\theequation}{S\arabic{equation}}
% Give SI destinations unique PDF anchor names
\renewcommand{\theHsection}{supp.\arabic{section}}
\renewcommand{\theHfigure}{supp.\arabic{figure}}
\renewcommand{\theHtable}{supp.\arabic{table}}
\renewcommand{\theHequation}{supp.\arabic{equation}}
\renewcommand{\theHsubsection}{supp.\arabic{section}.\arabic{subsection}}
\renewcommand{\theHsubsubsection}{supp.\arabic{section}.\arabic{subsection}.\arabic{subsubsection}}

\begin{bibunit}[unsrt]

\begin{center}
{\Large\bfseries Supplementary Information\par}
\vspace{0.5em}

{\fontsize{12}{15}\selectfont
Cosm: orchestrated collective dynamics give rise to rapid computational search at the solution frontier\par}

\vspace{0.75em}

Kenneth M. Zick, Nikhil Shukla, Alexander Marakov
\end{center}

\vspace{1em}

\section{Experimental details -- scaling experiment}
\label{sec: scaling exp details}
To match the prior Hou et al. study, scaling characterization was limited to the five largest sizes shown in Fig. 1 of Ref.~\cite{hou2025}, corresponding to sizes of $L\in\{12,14,16,18,20\}$. At each size, $N_\mathrm{sweeps}$ was tuned, while other parameters were held fixed: initial step size $\alpha_0$ = 25 degrees; $r_\mathrm{DWT}$ = 1.0; $T_\mathrm{DWT}$ = 3 sweeps. Thus, the scaling characterization allows size-dependent optimization of run length while retaining common values of the other dynamical parameters across the relatively narrow fitted range of $L=12$–$20$. The tuned values of $N_\mathrm{sweeps}$ were 300, 500, 700, 800, and 1200 for $L\in\{12,14,16,18,20\}$, respectively. An instance with estimated median $P_\mathrm{s}$ was identified using a MATLAB 2024b-based implementation of Cosm; the instance IDs were 27, 41, 64, 2, and 8 for $L\in\{12,14,16,18,20\}$, respectively. The execution times of the five median instances were then measured using an unoptimized Cosm C implementation running on a Xeon Gold 6544Y. (Note that while MATLAB was used for parameter tuning and identifying median instances, all reported timing results are based on the C implementation.) In this experiment, floating-point variables in C used double precision. Since these are square lattice instances, Cosm uses a standard 4-color edge coloring that is trivial to assign; the time required to assign colors in preprocessing is not included in execution time measurements. Average execution times per trial were measured by performing 10,000 trials of each median instance. The trials were performed sequentially (no parallel replicas); use of sequential trials and a single job matches Ref.~\cite{hou2025}, which used a single CPU core. The observed $P_\mathrm{s}$ and average execution time were used to calculate the median $TTS$. Following Hou et al., we fit $log_{10}TTS$ data to $a+bL$ and obtained a slope representing the scaling exponent $b$.

% ==================================================
\section{Experimental details -- Gset experiments}
\label{sec:exp_details_Gset}

All 2D lattice instances used a 4-color edge coloring with the same regular pattern. For non-lattice graph instances, the deterministic greedy edge-coloring heuristic (Sec.~\ref{sec:pseudocode}) was used. For each undirected edge, the smaller node index was placed first, and the resulting edge list was sorted first by the first endpoint and then by the second endpoint. This matches the ordering produced by MATLAB's \texttt{graph(src, dst)} function. A C implementation of the edge-coloring heuristic was used for characterizing preprocessing time; it uses the same fixed ordering and therefore generates the same coloring. The number of edges in each edge color partition for G60/61/70 is shown in Supplementary Table \ref{tab:num_edges_by_color}. G64 is omitted from the table because its broad degree distribution results in 589 edge-color partitions. Alternative edge-coloring heuristics are possible that use additional colors or balance the number of edges per color.  Although a minimum coloring is not strictly necessary, the edge-coloring heuristic did generate a minimum coloring for each of the four non-lattice graphs, with the number of colors equal to the maximum vertex degree. For all experiments, we employed step size schedules that started from an initial value $\alpha_0$ and ramped down linearly to 0. The step size was updated once per sweep and stayed constant within a sweep. At the end of each trial, the continuous circular variables were binarized by evaluating 100 evenly spaced bisector orientations of the unit circle (1.8-degree spacing over a semicircle) using the Max-Cut objective and taking the best solution. We selected a value of 100 as a conservative value that empirically was sufficient to obtain the reported solution quality. Parameter settings common to all Gset experiments are shown in Supplementary Table \ref{tab:common_parameter_settings}.

The G72/77/81 and ablation experiments used an unoptimized Cosm C implementation with single-precision floating point, compiled with GNU gcc 8.5.0 (-O3 -ffast-math option). To evaluate solution quality on G72/77/81, we performed 1024 trials per instance. The highest cut attained for each instance was compared with previously reported solutions and with the optimum, which we certified using Gurobi 12.0.3 with default solver settings. The Cosm solver parameter settings are given in Supplementary Table~\ref{tab:Gset_nonTTT_parameter_settings}.

\begin{table}[h]
\centering
\caption{Number of edges per color using greedy edge-coloring heuristic.}
\label{tab:num_edges_by_color}
% Resize implies we force it to textwidth, ensuring it fits perfectly.
%\resizebox{\columnwidth}{!}{%
\begin{tabular}{c l}
\toprule
%\shortstack{\textbf{Gset}\\\textbf{Instance}} &
  \textbf{Gset instance} &
  \textbf{Number of edges per color ($|E_1|$, $|E_2|$, ... $|E_C|$)} \\
\midrule
G60 & 3036, 2862, 2608, 2331, 1949, 1561, 1150, 777, 467, 251, 103, 37, 12, 4 \\
G61 & 3036, 2862, 2608, 2331, 1949, 1561, 1150, 777, 467, 251, 103, 37, 12, 4 \\
G70 & 3440, 2747, 1925, 1135, 523, 170, 44, 11, 4 \\
\bottomrule
\end{tabular}
\end{table}

\begin{table}[h]
\centering
\caption{Common parameter settings for all Gset experiments.}
\label{tab:common_parameter_settings}
%\resizebox{0.7\columnwidth}{!}{
\begin{tabular}{l c}
\toprule
  \textbf{Parameter} &
  \textbf{Value} \\
\midrule
Final step size & 0 degrees \\
Size of each DWT window & 90 degrees \\
Number of candidate bisectors for binarization & 100 \\
\bottomrule
\end{tabular}
\end{table}

\begin{table}[htbp] %[H]
\centering
\caption{Parameter settings for Gset solution quality and ablation experiments.}
\label{tab:Gset_nonTTT_parameter_settings}
%\resizebox{0.88\columnwidth}{!}{
\begin{tabular}{l c c c c}
\toprule
  \textbf{Parameter} &
  \textbf{G70} &
  \textbf{G72} &
  \textbf{G77} &
  \textbf{G81} \\
\midrule
Initial step size $\alpha_0$ (degrees) & 14 & 14 & 14 & 14 \\
$r_\mathrm{DWT}$ & 1.0 & 0.5 & 0.5 & 0.5\\
DWT period $T_\mathrm{DWT}$ & 6 & 3 & 3 & 3 \\
$N_\mathrm{sweeps}$ & $1.5\times10^6$ & $10^6$ & $10^6$ & $2\times10^6$ \\
\bottomrule
\end{tabular}
\end{table}

All Gset $TTT$ timing measurements used an optimized Cosm C implementation with 16-bit unsigned integers, a branchless inner loop, and 32 parallel replicas matched to AVX-512 vector lanes. It was compiled with GNU gcc 8.5.0 (-O3 -march=native -flto -DNDEBUG). Within each 32-replica SIMD batch, the same randomly generated DWT reference angle $\theta_{\rm ref}$ was applied to all replicas at a given perturbation event, while each replica maintained an independently initialized state. To assess whether this shared randomness induced appreciable dependence among replica outcomes, we compared the observed distribution of target successes per SIMD batch with the binomial distribution predicted under independent outcomes. Close agreement was observed for all four Gset instances used for $TTT$ characterization, providing empirical support for the independence approximation used in equation (\ref{eq: P_s,wave}). For all Gset TTT experiments, DWT rotations used a single fixed direction rather than randomly selecting the rotation direction at each perturbation. Success probability and execution time were characterized separately. For each benchmark instance, the single-replica success probability $P_\mathrm{s}$ was estimated from 128,000 trials across multiple jobs, with each job initialized with a distinct seed. In contrast, all wall-clock timing measurements used to calculate $TTT$ were performed on the fully loaded $2\times$ Xeon Gold 6544Y system. Wave execution time $t_{\mathrm{wave}}$ was measured with 32 concurrent SIMD jobs, corresponding to 1024 replicas. The measured single-replica $P_\mathrm{s}$ was used to calculate $P_{s,\mathrm{wave}}$ according to equation (\ref{eq: P_s,wave}), and this probability was combined with the measured wave time (averaged over 200 waves for each instance) to calculate $TTT_{99}$. Solver parameter settings are given in Supplementary Table~\ref{tab:Gset_TTT_parameter_settings}.

\begin{table}[h]
\centering
\caption{Parameter settings for Gset $TTT$ experiments.}
\label{tab:Gset_TTT_parameter_settings}
%\resizebox{0.88\columnwidth}{!}{
\begin{tabular}{l c c c c}
\toprule
  \textbf{Parameter} &
  \textbf{G60} &
  \textbf{G61} &
  \textbf{G64} &
  \textbf{G70} \\
\midrule
Initial step size $\alpha_0$ (degrees) & 14 & 14 & 7 & 14 \\
$r_\mathrm{DWT}$ & 1.5 & 1.5 & 3.5 & 1.0 \\
DWT period $T_\mathrm{DWT}$ & 3 & 3 & 3 & 3 \\
$N_\mathrm{sweeps}$ & 230,000 & 370,000 & $10^6$ & {80,000} \\
\bottomrule
\end{tabular}
\end{table}

% ==================================================
\section{Comparison with previously reported solution quality}
\label{sec: prior_highest_cuts}

This section provides additional context for the solution quality obtained by Cosm on the large Gset instances considered in Section \ref{sec:results}. Supplementary Table~\ref{tab:cut_comparisons} compares the highest cuts attained by Cosm with selected previously reported results, including both heuristic and exact-solver calculations. Because the cited studies differ substantially in computational resources, run lengths, parameter-tuning procedures, and reporting protocols, the table is intended as a comparison of attained solution quality rather than computational performance, and not as a comprehensive historical record. Ikuta et al. reported CPLEX calculations that certified the optimum for G72 and attained the optimal cut for G77 without establishing its optimality~\cite{ikuta2015}.

\begin{table}[h]
\centering
\caption{Solution quality of Cosm and selected prior works on Gset instances. Bold values indicate the highest cuts shown for each instance. Values in parentheses indicate higher cuts reported during calibration rather than standard testing. Dashes indicate results not reported in the cited studies.}
\label{tab:cut_comparisons}
% Resize implies we force it to textwidth, ensuring it fits perfectly.
%\resizebox{\textwidth}{!}{%
\setlength{\tabcolsep}{5pt}
\renewcommand{\arraystretch}{1.5}
\begin{tabular}{l c c c c c c c}
\toprule
\textbf{Solver approach} & \textbf{G60} & \textbf{G61} & \textbf{G64} & \textbf{G70} & \textbf{G72} & \textbf{G77} & \textbf{G81} \\
\midrule
Cosm (this work) & \textbf{14190} & \textbf{5799} & \textbf{8753} & \textbf{9595} & \textbf{7008} & \textbf{9940} & \textbf{14060} \\
\addlinespace[5pt]
Fujitsu DAU (2020)~\cite{matsubara2020} & \textbf{14190} & \textbf{5799} & 8751 & --- & --- & --- & --- \\
CPLEX (2015)~\cite{ikuta2015} & 13589 & 5237 & 7340 & --- & \textbf{7008} & \textbf{9940} & 14054 \\
GESPR (2015)~\cite{shylo2015} & 14188 & 5796 & 8751 & 9591 & 7006 & 9938 & 14048 \\
Large population island (2024)~\cite{goudet2024} & \textbf{14190} & \makecell{5798\\(\textbf{5799})} & 8752 & \makecell{9594\\(\textbf{9595})} & 7004 & 9926 & 14030 \\
MOH (2017)~\cite{ma2017} & \textbf{14190} & 5798 & 8747 & 9544 & 6998 & 9928 & 14036 \\
Breakout Local Search (2013)~\cite{benlic2013} & 14176 & 5789 & 8735 & 9541 & 6998 & 9926 & 14030 \\
\makecell[l]{Simulated bifurcation\\machine (2021)~\cite{goto2021}} & 14186 & 5796 & 8739 & 9578 & 6982 & 9904 & 13992 \\
\makecell[l]{Fixstars Amplify\\Annealing Engine (2025)~\cite{fixstarsamplifydoc2025,liu2025}} & 14175 & 5796 & 8744 & 9582 & 6970 & 9856 & --- \\
Simulated annealing (2026)~\cite{vodeb2026} & 14114 & 5788 & 8728 & 9546 & 6984 & --- & --- \\
D-Wave Hybrid (2026)~\cite{vodeb2026} & 14142 & 5750 & 8665 & 9516 & 6914 & --- & --- \\
CirCut (2002)~\cite{burer2002} & 14081 & 5690 & 8575 & 9529 & 6820 & 9670 & 13662 \\ 
\bottomrule
\end{tabular}%
\end{table}

% ==================================================
\FloatBarrier
\section{Cosm energy-to-target for Gset instances G60/61/64/70}
\label{sec:ETT}

We calculate Cosm's CPU and DRAM energy-to-target using
\begin{equation}
\label{eq:ETT}
\begin{aligned}
ETT_{99,\mathrm{CPU}(target)} &= POW_{\mathrm{CPU}} \cdot TTT_{99}(target), \\
ETT_{99,\mathrm{DRAM}(target)} &= POW_{\mathrm{DRAM}} \cdot TTT_{99}(target)
\end{aligned}
\end{equation}
where $POW$ is the measured average power consumption during 1024-replica waves (32 parallel batches and 32 SIMD replicas per batch) for each instance, and the $TTT_{99}$ values were characterized as described in Section~\ref{sec:performance_metrics}. The energies consumed by the two CPU packages and two DRAM subsystems of the 2× Xeon Gold 6544Y system during 1024-trial waves were measured using Intel Running Average Power Limit (RAPL). $POW_{\mathrm{CPU}}$ represents the total power across the two CPU packages. Power was calculated from the measured energy consumption and wall clock time of each wave. The mean and standard deviation of the power samples were calculated over 200 waves. Note that the observed power of the fully-loaded pair of CPUs is roughly consistent with the thermal design power specification ($2\times270 \mathrm{W}=540 \mathrm{W}$).  

\begin{table}[htbp]
\centering
\caption{Cosm energy-to-target for Gset instances G60/61/64/70.}
\label{tab: ETT}
\setlength{\tabcolsep}{4pt}
\begin{tabular}{c r r l l r r}
\toprule
  \makecell{\textbf{Gset}\\ \textbf{instance}} & 
  \makecell{\textbf{Target}\\ \textbf{solution}} & 
  \makecell{$\boldsymbol{TTT_{99}}$\\ \textbf{(s)}} &
  \makecell{$\boldsymbol{POW_{\mathrm{CPU}}}$\\ \textbf{(W)}} &
  \makecell{$\boldsymbol{POW_{\mathrm{DRAM}}}$\\ \textbf{(W)}} &
  \makecell{$\boldsymbol{ETT_{99,\mathrm{CPU}}}$\\ \textbf{(kJ)}} &
  \makecell{$\boldsymbol{ETT_{99,\mathrm{DRAM}}}$\\ \textbf{(kJ)}} \\
\midrule
G60 & 14190 & 9.46 & 527.56$\pm$10.89 & 17.47$\pm$0.04 & 4.99 & 0.17 \\
G61 & 5799 & 15.05 & 529.85$\pm$6.99 & 17.46$\pm$0.04 & 7.97 & 0.26 \\
G64 & 8753 & 205.8 & 528.41$\pm$4.97 & 17.47$\pm$0.04 & 108.75 & 3.60 \\
G70 & 9595 & 2.43 & 508.49$\pm$32.64 & 17.49$\pm$0.11 & 1.24 & 0.04 \\
\bottomrule
\end{tabular}
\end{table}

% ==================================================
\section{New best-known solution to Gset instance G64}
\label{sec:G64_8753_bitstring}

Cosm discovered a new best-known cut of 8753 on Gset instance G64. The Ising energy of this solution is $-16979$. The previous best-known cut was 8752~\cite{goudet2024}. A Cosm solution bitstring representing a cut of 8753 is provided below. The solution can be verified by expanding the hexadecimal string into a 7,000-bit configuration representing variables 1 through 7,000, mapping the \{0,1\} values to $\{-1,+1\}$ using either convention, then evaluating the Max-Cut objective in equation (\ref{eq:maxcut}) using the mapped values and the G64 problem instance defined by Ref.~\cite{gset}. 

\vspace{1em}
{\ttfamily\small\noindent\seqsplit{1aa3afee1e48838092d9d58744bafd15bb3361383977dfaacd44aac9e6bc0312e6a6b04c83dcb6c77031083fa888add9f1cb1cd587312c3aeb32a7b77928cc035647a6dc719986a7dd2f247b33b7f2e4815fe4a774517fcbb26e98e9ba2717c169bb35f3f942423c605d4dcf52bb23fcfdc180e45afa6e892e688046b93919ee409fee0553c740112136c250d9ca6a1fde140d9d22f00f57792a2cb478faf42a960ccc2d1a4a02f7cc2b4e2414bfb8b013b95ae588d8244677079c487ba9b37f64e071a3c0b9c821d1ae79a8477d0aee81239245d095b1706da6e178ea395357587a1639fbdd50aa4721c229145d49c23a361cf4442a57dac31a12630d1cad0e2147ad7e92b55ff1835e20004765cb388d837acd519eabd085b5c05122926a676a5d3114aac80e983631eb6d4400bc6206489dfbb8be1d6c45a476496a129753520feedd68c8dd24997f0121bb3923d94ff12071b3aba9b74453032730da20499208a6a681686b7e7cd4756afc6301dabe588611e561a4d9b4fe052fc43e07a4a957ff793db88a2213db759816da3c00cd1dac324b118afc3dac2656b05b41d8d7a42abb5b3c91ad7024d991ee2c4595efc4b1972e70fd0f97a19c9c4e9e1ca4d60c3504ece75aeb49b10c38f80e6f51ad14c38d3614b2d9d2a94945ad6d635000584e651f69eecf9aca10ffadbf83879316677a081e2e548e7d3cfb440e9333a0325102c1a3c474627238e7425b2fcc5622c7fe8e9303d9b91f209308defe7311c3f04ce0e3378bd348ae842e77f3acdefc1126d89ccf3d3733433441f0973f0681bf89f0f8d939844fbd73105fec876bd55123fcf6e989f455f2a82e86c840560ad3e8dca12614b777ef7731822bae3c39b7b4ff5f916e7a28b5efa871570063474db3b45ecf82f917a5e5feaf5e8dc3bc28472e9c2981df7828625eaa59814cd1bc8b95abee7de10860ddca20f50cf214d59ed9cf6889fb0dcf12307fd12aefb0f3c9816962351b2832cda38bafe5a8034981e6a6ed4a60a97c142cda59267af1ec0f9b01c91e7369f99f7f933eafbad3abe392512aa99fd8fdab42e1db3783ee7f5e9110095a6ba2162f01217660bb5d6ab87b78490b4d9144e1dbe65c4c062aa1ca2e55a8c5025e30abeaf9c4ce366c47ea7c1435e3bc5dfd5110c5e9f723da5e607b99ee74b53d3788dc732a28fd5093187cb2bed3432de2997bd48f2bd604bebbc3654f0ad49a43}}

\renewcommand{\refname}{Supplementary References}
\putbib[Cosm_references]

\end{bibunit}

\end{document}